\documentclass[twocolumn]{aastex631}

\newcommand\swift{\textit{Swift}}

\received{25 July 2024}

\submitjournal{ApJ}

\begin{document}

\title{The X-ray Luminous Type Ibn SN 2022ablq: Estimates of Pre-explosion Mass Loss and Constraints on Precursor Emission}

\correspondingauthor{Craig Pellegrino}
\email{cmp5cr@virginia.edu}

\author[0000-0002-7472-1279]{C. Pellegrino}
\affil{Department of Astronomy, University of Virginia, Charlottesville, VA 22904, USA}

\author[0000-0001-7132-0333]{M. Modjaz}
\affil{Department of Astronomy, University of Virginia, Charlottesville, VA 22904, USA}

\author[0000-0002-8215-5019]{Y. Takei}
\affil{Yukawa Institute for Theoretical Physics, Kyoto University, Kitashirakawa-Oiwake-cho, Sakyo-ku, Kyoto, Kyoto 606-8502, Japan}
\affil{Research Center for the Early Universe (RESCEU), School of Science, The University of Tokyo, Bunkyo, Tokyo 113-0033, Japan}
\affil{Astrophysical Big Bang Laboratory, RIKEN, 2-1 Hirosawa, Wako, Saitama 351-0198, Japan}

\author[0000-0002-6347-3089]{D. Tsuna}
\affil{TAPIR, Mailcode 350-17, California Institute of Technology, Pasadena, CA 91125, USA}
\affil{Research Center for the Early Universe (RESCEU), School of Science, The University of Tokyo, Bunkyo, Tokyo 113-0033, Japan}

\author[0000-0001-9570-0584]{M. Newsome}
\affil{{Las Cumbres Observatory, 6740 Cortona Drive, Suite 102, Goleta, CA 93117-5575, USA}}
\affil{{Department of Physics, University of California, Santa Barbara, CA 93106-9530, USA}}

\author[0000-0001-9227-8349]{T. Pritchard}
\affil{NASA Goddard Space Flight Center, 8800 Greenbelt Road, Greenbelt, MD 20771, USA}

\author[0009-0004-7268-7283]{R. Baer-Way}
\affil{Department of Astronomy, University of Virginia, Charlottesville, VA 22904, USA}

\author[0000-0002-4924-444X]{K.~A. Bostroem}\altaffiliation{LSST-DA Catalyst Fellow}
\affil{Steward Observatory, University of Arizona, 933 North Cherry Avenue, Tucson, AZ 85721-0065, USA}

\author[0000-0002-0844-6563]{P. Chandra}
\affil{National Radio Astronomy Observatory, 520 Edgemont Rd, Charlottesville VA 22903, USA}

\author[0000-0002-0326-6715]{P. Charalampopoulos}
\affil{Department of Physics and Astronomy, University of Turku, FI- 20014 Turku, Finland}

\author[0000-0002-7937-6371]{Y. Dong}
\affil{Department of Physics and Astronomy, University of California, 1 Shields Avenue, Davis, CA 95616-5270, USA}

\author[0000-0003-4914-5625]{J. Farah}
\affil{{Las Cumbres Observatory, 6740 Cortona Drive, Suite 102, Goleta, CA 93117-5575, USA}}
\affil{{Department of Physics, University of California, Santa Barbara, CA 93106-9530, USA}}

\author[0000-0003-4253-656X]{D.~A. Howell}
\affil{{Las Cumbres Observatory, 6740 Cortona Drive, Suite 102, Goleta, CA 93117-5575, USA}}
\affil{{Department of Physics, University of California, Santa Barbara, CA 93106-9530, USA}}

\author[0000-0001-5807-7893]{C. McCully}
\affil{{Las Cumbres Observatory, 6740 Cortona Drive, Suite 102, Goleta, CA 93117-5575, USA}}

\author[0000-0002-1856-9225]{S. Mohamed}
\affil{Department of Astronomy, University of Virginia, Charlottesville, VA 22904, USA}

\author[0000-0003-0209-9246]{E. Padilla Gonzalez}
\affil{{Las Cumbres Observatory, 6740 Cortona Drive, Suite 102, Goleta, CA 93117-5575, USA}}
\affil{{Department of Physics, University of California, Santa Barbara, CA 93106-9530, USA}}

\author[0000-0003-0794-5982]{G. Terreran}
\affil{{Las Cumbres Observatory, 6740 Cortona Drive, Suite 102, Goleta, CA 93117-5575, USA}}
\affil{{Department of Physics, University of California, Santa Barbara, CA 93106-9530, USA}}

\shortauthors{Pellegrino et al.}
\shorttitle{X-ray Luminous SN 2022ablq}

\begin{abstract}

Type Ibn supernovae (SNe Ibn) are rare stellar explosions powered primarily by interaction between the SN ejecta and H-poor, He-rich material lost by their progenitor stars. Multi-wavelength observations, particularly in the X-rays, of SNe Ibn constrain their poorly-understood progenitor channels and mass-loss mechanisms. Here we present \textit{Swift} X-ray, ultraviolet, and ground-based optical observations of the Type Ibn SN\,2022ablq–only the second SN Ibn with X-ray detections to date. While similar to the prototypical Type Ibn SN\,2006jc in the optical, SN\,2022ablq is roughly an order of magnitude more luminous in the X-rays, reaching unabsorbed luminosities $L_X$ $\sim$ 3$\times$10$^{40}$ erg s$^{-1}$ between 0.2 -- 10 keV. From these X-ray observations we infer time-varying mass-loss rates between 0.05 -- 0.5 $M_\odot$ yr$^{-1}$ peaking 0.5 -- 2 yr before explosion. This complex mass-loss history and circumstellar environment disfavor steady-state winds as the primary progenitor mass-loss mechanism. We also search for precursor emission from alternative mass-loss mechanisms, such as eruptive outbursts, in forced photometry during the two years before explosion. We find no statistically significant detections brighter than M $\approx$ -14\textemdash too shallow to rule out precursor events similar to those observed for other SNe Ibn. Finally, numerical models of the explosion of a $\sim$15 $M_\odot$ helium star that undergoes an eruptive outburst $\approx$1.8 years before explosion are consistent with the observed bolometric light curve. We conclude that our observations disfavor a Wolf-Rayet star progenitor losing He-rich material via stellar winds and instead favor lower-mass progenitor models, including Roche-lobe overflow in helium stars with compact binary companions or stars that undergo eruptive outbursts during late-stage nucleosynthesis stages.

\end{abstract}


\section{Introduction} \label{sec:intro}

While the majority of stars with zero-age main sequence (ZAMS) masses $\gtrsim$ 8 $M_\odot$ explode as H-rich core-collapse supernovae \citep[SNe; e.g.,][]{Janka2012,Perley2020}, $\approx$ 30\% produce SNe which lack H and sometimes He in their ejecta \citep{Shivvers2017a}. These stripped-envelope SNe (SESNe) are thought to be the explosions of stars that have lost their outer H- (and He-) rich material before exploding \citep{Filippenko1997,Modjaz2019}. However, the nature of these progenitor stars, including their initial masses, binarity, and mass-loss mechanisms, remain open questions. Although their mass loss is reminiscent of the strong stellar wind-driven mass loss of Wolf-Rayet (WR) stars\textemdash the terminal stages of stars with masses M$_{\text{ZAMS}}$ $\gtrsim$ 25 $M_\odot$ \citep{Crowther2007}\textemdash it is becoming clear that WR stars cannot be the sole progenitors of SESNe on the basis of pre-explosion imaging, rates, and ejecta properties \citep[e.g.,][]{Drout2011,Bersten2014,Fremling2014,Taddia2018,Prentice2019,Kilpatrick2021}. Instead, a fraction (perhaps even the majority) of SESNe may come from lower-mass stars stripped from interaction with a binary companion \citep[e.g.,][]{Sana2012,Yoon2017,Dessart2020}.

A small subset \citep[$\approx$ 10\%;][]{Perley2020} of SESNe show spectroscopic signatures of interaction between the SN ejecta and pre-existing circumstellar material (CSM) lost by their progenitors. If the CSM is H-poor but He-rich, these SNe are classified as Type Ibn SNe (SNe Ibn). SNe Ibn are observationally rare, with $\sim$50 discovered to date \citep{Pastorello2008,Hosseinzadeh2017}. This in part is due to their rapidly-evolving light curves, which rise and fall in brightness faster than most other SNe subclasses\textemdash many SNe Ibn have characteristic timescales $t_{1/2}$ $\lesssim$ 12 days, where $t_{1/2}$ measures the rest-frame time the SN luminosity is above half its maximum value \citep{Ho2023}. This rapid evolution is driven by strong interaction between the ejecta and dense, confined CSM, which efficiently converts the kinetic energy of the SN ejecta into radiation. The collision between these media produces shocks which propagate forward through the CSM and backward through the ejecta, sweeping up the traversed material. This material then reprocesses X-ray photons from the shocks into the ultraviolet (UV) and optical, creating narrow ($\sim$ 1000 km s$^{-1}$) spectroscopic emission lines \citep[see e.g.,][for a review]{Chevalier1982,Chevalier1994}.

Recent observations of core-collapse SNe have revealed that a significant fraction \citep[$\gtrsim$ 30\%;][]{Bruch2023} of massive stars undergo periods of enhanced mass loss ($\gtrsim$ 10$^{-2}$ M$_\odot$ yr$^{-1}$) immediately preceding explosion, as demonstrated by the nearby, well-studied Type II SN\,2023ixf \citep[e.g.,][]{Bostroem2023,Hiramatsu2023,JacobsonGalan2023,Chandra2024}. However, the exact nature of these mass-loss mechanisms remains unclear. Despite their rarity, SNe Ibn offer one of the most promising avenues to study the mass-loss mechanisms in the final years of massive stars' lives. Their photometric and spectroscopic evidence for strong interaction immediately after explosion suggests that their progenitors must have undergone significant, elevated mass loss during their final days to years. Proposed progenitor channels include the core-collapse of WR stars that have either lost their outer envelopes through strong stellar winds or eruptive outbursts similar to those seen in Luminous Blue Variable (LBV) stars \citep{Pastorello2007,Foley2007}, or lower-mass stars that undergo enhanced mass loss either owing to Roche-lobe overflow or to unstable shell burning \citep[e.g.,][]{Wu2022}. Several studies of SNe Ibn have indirectly inferred their mass loss histories and progenitor properties through light curve modeling, which allows for rough estimates of their CSM and ejecta properties. Many of these results reveal a surprising uniformity in their properties, with CSM masses $M_{\text{CSM}}$ $\lesssim$ 1 $M_\odot$, low ejecta masses $M_{\text{ej}}$ $\sim$ 1-3 $M_\odot$, and low synthesized $^{56}$Ni mass estimates or upper limits $M_{\text{Ni}}$ $\lesssim$ 0.05 $M_\odot$ \citep[e.g.,][]{Clark2020,Karamehmetoglu2021,Pellegrino2022,Maeda2022,BenAmi2023,Dong2024}, although some notable exceptions exist \citep[e.g.,][]{Pastorello2015a,Kool2021,Wang2021}. 

Detections of pre-explosion emission at the SN position also probe the progenitor's mass loss history. For example, a short-lived precursor to the prototypical Type Ibn SN\,2006jc was observed more than two years before the terminal SN explosion \citep{Pastorello2007}. This precursor is inconsistent with steady state wind-driven mass loss \citep[e.g.,][]{Tominaga2008} and instead has been interpreted as an eruptive outburst of the progenitor star \citep[][see also \citealt{Tsuna2024a}]{Maund2016}. A similar duration outburst was recovered in pre-explosion images of the Type Ibn SN\,2019uo \citep{Strotjohann2021}. More recently, the nearby Type Ibn SN\,2023fyq displayed a remarkable, long-duration ($\gtrsim$ 3 year) pre-explosion light curve \citep{Brennan2024} which has been suggested to be evidence of a He star progenitor overflowing its Roche-lobe in a binary system \citep{Dong2024,Tsuna2024}.

X-ray observations are a more direct, though less commonly-used, tool to directly measure the mass-loss histories of SN progenitors. X-rays are emitted by the forward and reverse shocks that are produced by the ejecta-CSM interaction \citep{Chevalier1994}. The timescale of the X-ray evolution and its luminosity allows for direct measurements of the CSM mass and density over time, enabling one to map the circumstellar environment and infer the mass-loss rate during the final days to years of the progenitor stars' lives \citep[e.g.,][]{Immler2001,Tsuna2021,Margalit2022}. While, to date, SN\,2006jc is the only SN Ibn with a published X-ray light curve \citep{Immler2008}, its X-ray observations revealed a wealth of information about its circumstellar environment and progenitor's mass-loss history. Its X-ray light curve rose to a peak $\sim$100 days after explosion, which \citet{Immler2008} inferred to be evidence of the forward shock traversing a shell of material lost by its progenitor star two years prior\textemdash an independent verification of the observed optical precursor event. The X-ray evolution provided a lower estimate of the CSM mass $M_{\text{CSM}}$ $>$ 0.01 $M_\odot$.

Here we present observations of SN\,2022ablq, only the second SN Ibn with unambiguous X-ray detections. SN\,2022ablq offers the rare opportunity to directly probe the properties of an SN Ibn progenitor. At a distance of only 60 Mpc (Section \ref{sec:methods}), SN\,2022ablq is one of the closest SNe Ibn in recent years, with a large dataset of pre-explosion images probing the years before the progenitor's terminal explosion. Together with its X-ray detections, these data allow for an unparalleled look into the mass-loss history and circumstellar environment of an SN Ibn progenitor.

This paper is organized as follows. We detail the discovery and follow-up observations of SN\,2022ablq in Section \ref{sec:methods}. In Section \ref{sec:optical} we present its UV-optical light curves and spectra and affirm its classification as a Type Ibn SN. We  analyze its X-ray light curve evolution in Section \ref{sec:x-ray} and provide constraints on its pre-explosion emission in Section \ref{sec:preexplosiondata}. In Section \ref{sec:lcmodel} we model the post-explosion light curve with a numerical model that self-consistently reproduces the inferred mass-loss and CSM parameters. Finally, we discuss the implications of our results and conclude in Section \ref{sec:discussion}.

\section{Discovery and Data Description}\label{sec:methods}

SN\,2022ablq (also known as ASASSN-22nu and ATLAS22bmxy) was discovered by the All-Sky Automated Survey for Supernovae (ASAS-SN) on UT 2022-11-24 13:55:12.000 \citep[MJD 59907.58;][]{Stanek2022}. \citet{Fulton2022} submitted a classification report to the Transient Name Server (TNS) using a spectrum obtained on UT 2022-11-26 05:28:42 (MJD 59909.23), identifying SN\,2022ablq as a peculiar transient with spectroscopic similarities to both SNe Ibn as well as Tidal Disruption Events (TDEs). At the time of this classification, we began high-cadence photometric and spectroscopic observations of the transient with the Las Cumbres Observatory \citep[LCO;][]{Brown2013} telescope network as part of the Global Supernova Project. The first LCO photometric and spectroscopic observations were obtained 0.75 and 2.8 days after the classification report, respectively, at which point the transient was already around peak brightness. 

SN\,2022ablq exploded at R.A. 12$^{\text{h}}$13$^{\text{m}}$06$^{\text{s}}$.48 and decl. $+$17\textdegree05$^{\prime}$56$^{\prime\prime}$.2, at an angular separation of 0.36\arcsec{} from the nucleus of its host galaxy MRK 0762. The redshift of this galaxy, from an archival Sloan Digital Sky Survey (SDSS) spectrum, is $z =  0.0143$. Throughout this paper we adopt this redshift and corresponding luminosity distance $d_{\text{lum}} = 63$ Mpc, assuming a flat $\Lambda$CDM cosmology with $H_0 = 67.7$ km s$^{-1}$ Mpc$^{-1}$ \citep{Planck2020}. We assume a Galactic line-of-sight extinction toward SN\,2022ablq $E(B-V)_{\text{MW}}$ = 0.031 mag, using the dust map calibrations of \citet{Schlafly2011}. To estimate its host galaxy extinction given the lack of high-resolution spectra, we follow the procedure of \cite{Gangopadhyay2020} and compare the peak $g-r$ colors of SN\,2022ablq to that of SN\,2019uo, another well-observed SN Ibn around maximum light \citep{Gangopadhyay2020,Pellegrino2022}. Their colors are consistent if we assume a host galaxy extinction $E(B-V)_{\text{host}}$ = 0.18 mag (given $R_V$ = 3.1). While there is evidence for a spread in the optical colors of SNe Ibn around peak brightness \citep[e.g.,][]{BenAmi2023}, assuming this additional host extinction also brings the light curve of SN\,2022ablq in agreement with the typical behavior of its class (see Section \ref{subsec:lc}). Therefore, despite the inherent uncertainty in this method, we adopt a total extinction $E(B-V)_{\text{tot}}$ = 0.21 mag throughout this work. All photometry presented has been corrected for extinction.

\subsection{Optical Photometry}

\begin{deluxetable}{ccccc}[t]
\tablecaption{UV and Optical Photometry \label{tab:phot}}
\tablehead{
\colhead{MJD} & \colhead{Magnitude} & \colhead{Uncertainty} & \colhead{Filter} & \colhead{Source}}
\startdata
59919.43 & 13.02 & 0.07 & UVW2 & UVOT\\
59923.47 & 13.70 & 0.09 & UVW2 & UVOT\\
59927.45 & 14.30 & 0.11 & UVW2 & UVOT\\
59942.48 & 15.64 & 0.24 & UVW2 & UVOT\\
59946.69 & 15.89 & 0.29 & UVW2 & UVOT\\
59950.67 & 15.98 & 0.31 & UVW2 & UVOT\\
59919.34 & 12.87 & 0.07 & UVM2 & UVOT\\
\enddata
\tablenotetext{}{This table will be made available in its entirety in machine-readable format. A portion is shown here for reference.}
\end{deluxetable}

SN\,2022ablq was imaged with Sinistro cameras mounted on Las Cumbres Observatory 1.0 m telescopes in the \textit{UBgVri}-bands beginning 9 days after explosion. We reduced the data using the custom pipeline \texttt{lcogtsnpipe} \citep{Valenti2016}, which performs point-spread function fitting, calculates zero points and color terms \citep{Stetson1987}, and extracts photometric magnitudes. We calibrate \textit{UBV}-band photometry to Vega magnitudes using Landolt standard fields \citep{Landolt1992} and \textit{gri}-band photometry to AB magnitudes \citep{Smith2002} using Sloan Digital Sky Survey (SDSS) catalogs. Given the coincidence of SN\,2022ablq with its host galaxy, we performed template subtraction using the HOTPANTS \citep{Becker2015} algorithm and LCO template images that were obtained on 2023-06-12, after the SN had faded. 

We also requested forced photometry from the ATLAS \citep{Tonry2018,Smith2020} and ASAS-SN \citep{Shappee2014} forced photometry servers \citep{Shingles2021,Hart2023}. Post-explosion intranight flux measurements are combined via a weighted average in order to increase the signal-to-noise; we discuss our analysis of the pre-explosion data in Section \ref{sec:preexplosiondata}. Magnitudes are calculated from our weighted flux measurements in the ATLAS \textit{c} and \textit{o} bands and ASAS-SN \textit{g} band and calibrated to the AB magnitude system. Optical photometry data are presented in Table \ref{tab:phot}.

\subsection{Optical Spectra}

\begin{deluxetable*}{crlcc}[t!]
\tablecaption{Log of Spectroscopic Observations \label{tab:speclog}}
\tablehead{
\colhead{Date of Observation} & \colhead{Phase (days)} & \colhead{Facility/Instrument} & \colhead{Exposure Time (s)} & 
\colhead{Wavelength Range (\AA{})}}
\startdata
 2022-11-26 05:28:42 & -5.2 & LT/SPRAT & 1500 & 4000--8000 \\
 2022-12-02 06:22:07 & -0.2 & NOT/ALFOSC & 1800 & 3400--9700 \\
 2022-12-02 13:43:01 & 0.6 & LCO/FLOYDS & 1800 & 3500--10,000 \\
 2022-12-10 13:36:38 & 8.6 & LCO/FLOYDS & 1800 & 3500--10,000 \\
 2022-12-25 17:13:49 & 23.6 & LCO/FLOYDS & 1800 & 3500--10,000 \\
\enddata
\tablecomments{All spectra will be made publicly available on WiseRep \citep{Yaron2012}.}
\end{deluxetable*}

The Global Supernova Project obtained optical spectra of SN\,2022ablq between 2022-12-02 and 2022-12-25 using the FLOYDS spectrograph on LCO 2.0m telescopes in Haleakal$\bar{\text{a}}$ and Siding Springs. Spectra were reduced using a custom pipeline\footnote{https://github.com/svalenti/FLOYDS$\_$pipeline} which applies wavelength and flux calibration using a standard star observed on the same night. Given the strong contaminating host galaxy lines in the spectra, we use an archival SDSS spectrum of the underlying host galaxy to subtract the continuum and mask narrow host features (see Section \ref{subsec:spec}). We also include a 1800\,s optical spectrum obtained on 2022-12-02 using the Alhambra Faint Object Spectrograph and Camera mounted on the 2.56 m NOT located at La Palma, Spain (66-019; PI: P. Charalampopoulos). The spectrum was reduced using the spectroscopic data reduction pipeline \texttt{PyNOT}\footnote{\url{https://github.com/jkrogager/PyNOT}}. We used a 1.0 inch slit width and grism no. 4, covering the wavelength range $\sim 3200-9600$ \AA\, at resolution $\Delta\lambda/\lambda \approx 360$. The airmass during the observation was of the order of $\sim$ 1.3. Spectra information is given in Table \ref{tab:speclog}.

\subsection{\swift{} UVOT}

The Neils Gehrels \swift{} observatory \citep{Gehrels2004} obtained UV and optical photometry of SN\,2022ablq with the onboard Ultraviolet and Optical Telescope \citep[UVOT;][]{Roming2005} beginning 2022-12-06, 16 days after explosion (Proposer: Roy). We processed these data using an implementation of the Swift Optical/Ultraviolet Supernova Archive (SOUSA) pipeline \citep{Brown2014} with up-to-date calibration files and zeropoints \citep{Breeveld2011}. SN flux was extracted using a 3\arcsec{} aperture centered on the SN position. Background subtraction to mitigate host galaxy contamination was performed by subtracting the flux found in the same aperture from template images in each UVOT filter obtained on 2023-03-16, well after the SN had faded. All \swift{} photometry are calibrated to Vega magnitudes and given in Table \ref{tab:phot}.

\subsection{\swift{} XRT}\label{subsec:swiftxrt}

SN\,2022ablq was observed with the \swift{} X-ray Telescope (XRT) in Photon Counting mode simultaneously with the UVOT for a total of 48.2 ks. We processed the raw images using the \texttt{HEASoft} \citep{HEAsoft} version 6.32 routine \texttt{xrtpipeline}, which performs bias subtraction, screens bad pixels, and applies gain corrections to produce calibrated data products. Additionally, using \texttt{XSELECT}\footnote{https://heasarc.gsfc.nasa.gov/ftools/xselect/} we stacked the outputted science-ready images from \texttt{xrtpipeline} to increase the signal-to-noise. As SN\,2022ablq exploded near its host galaxy's nucleus, we carefully extract X-ray counts at the optical position of the SN using a circular aperture with a conservative radius of 3 pixels (7.2$\arcsec$). Given the low resolution of \swift{} XRT, this aperture encompasses less than half of the full-width at half-maximum (FWHM) of typical XRT images but also mitigates contaminating X-ray photons from the nearby host galaxy center \citep[e.g.,][]{Dwarkadas2016}. This aperture size is also consistent with similar analyses of other SN X-ray light curves \citep{Immler2008,Ofek2013}. Background counts were extracted using an annulus, centered at the SN position, with an inner radius of 142$\arcsec$ and outer radius of 260$\arcsec$, again chosen to mitigate the X-ray flux from the host galaxy. 

\section{Photometric and Spectroscopic Analysis}\label{sec:optical}

In this section we detail our analysis of the optical photometric and spectroscopic properties of SN\,2022ablq, as well as those of its host galaxy.

\begin{figure*}
    \centering
    \includegraphics[width=0.9\textwidth]{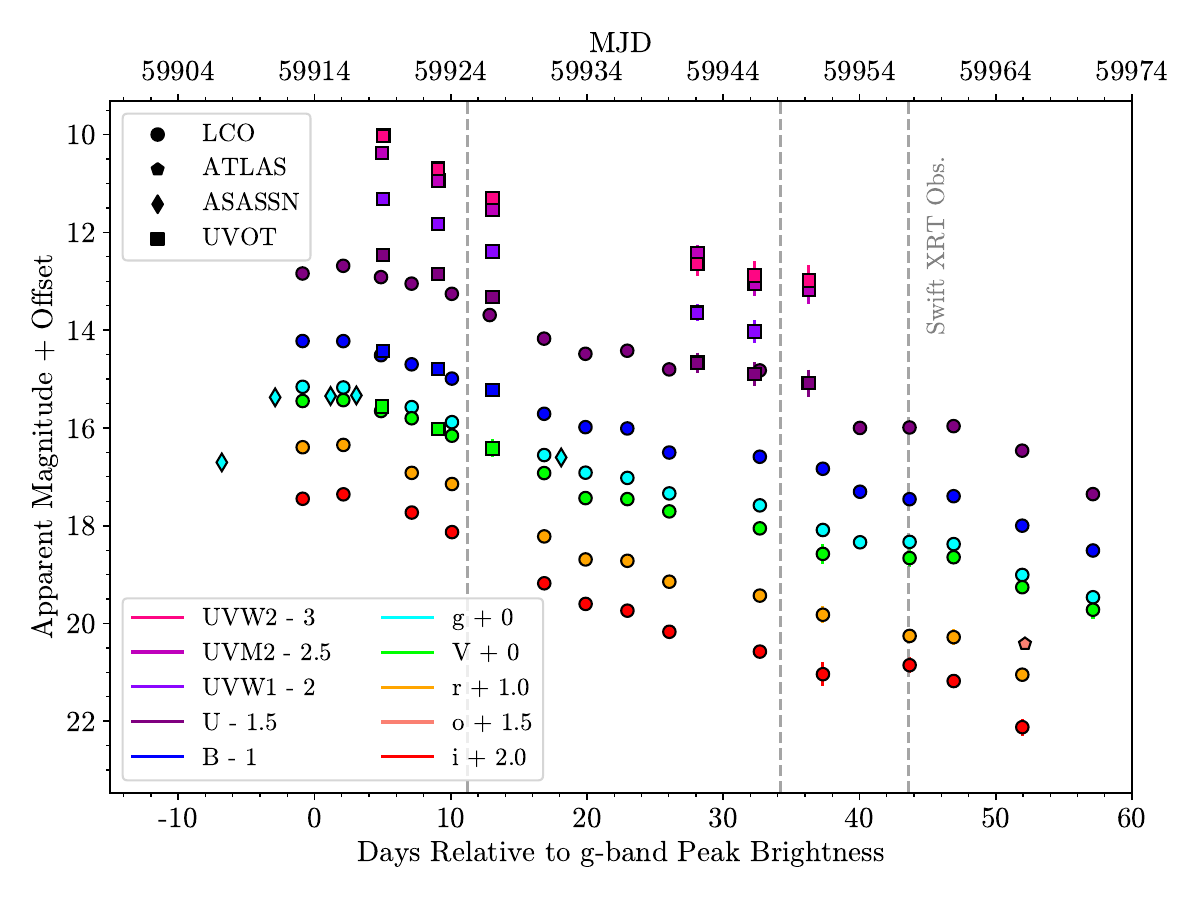}
    \caption{The UV and optical light curves of SN\,2022ablq from LCO (circles), ATLAS (pentagons), ASAS-SN (diamonds), and UVOT (squares). Phases are given relative to peak \textit{g}-band brightness. Offsets are added to each light curve for clarity. The approximate phases of the first three coadded \swift{} XRT observations are shown by a dashed vertical line\textemdash two additional XRT epochs were obtained after SN\,2022ablq fell below the detection threshold of our UV and optical observations, and are not displayed here.}
    \label{fig:lc}
\end{figure*}

\subsection{Ultraviolet - Optical Light Curves}\label{subsec:lc}

Figure \ref{fig:lc} shows the ultraviolet and optical light curves of SN\,2022ablq from \swift{}, LCO, ATLAS, and ASAS-SN. All data have been template subtracted and corrected for both Galactic and host galaxy extinction. \swift{} and LCO observations began around the time of the transient's peak brightness. The light curve shows luminous, rapidly-evolving emission at these phases, particularly in the UV bands. This evolution is similar to that of other SNe Ibn \citep[e.g.,][]{Hosseinzadeh2017}, as strong circumstellar interaction drives luminous and blue light curves. 

To estimate the time of peak brightness, we utilize the combined LCO and ASAS-SN \textit{g}-band light curves, as this is the only filter with pre-maximum coverage. We follow \citet{Bianco2014} and use a Monte Carlo routine to fit a third-order polynomial to the \textit{g}-band data around the observed time of peak brightness. In each iteration the phase range to fit is randomly varied to sample different light curve points and the fitted peak brightness and peak phase are recorded. We report the average and standard deviation of these values as our estimated \textit{g}-band peak brightness $m_{\text{peak}}$ =  15.20 $\pm$ 0.02 and time of peak $t_{\text{peak}}$ = MJD 59914.08 $\pm$ 1.04. We also attempt to constrain the time of explosion; however, SN\,2022ablq lacks high-cadence observations around the time of its first detection. We therefore conservatively estimate the explosion epoch as halfway between the last ASAS-SN nondetection and first detection, $t_{\text{exp}}$ = MJD 59903.58 $\pm$ 3.99. For a more detailed treatment of the pre-explosion data, see Section \ref{sec:preexplosiondata}.

\begin{figure*}
    \centering
    \includegraphics[width=0.45\textwidth]{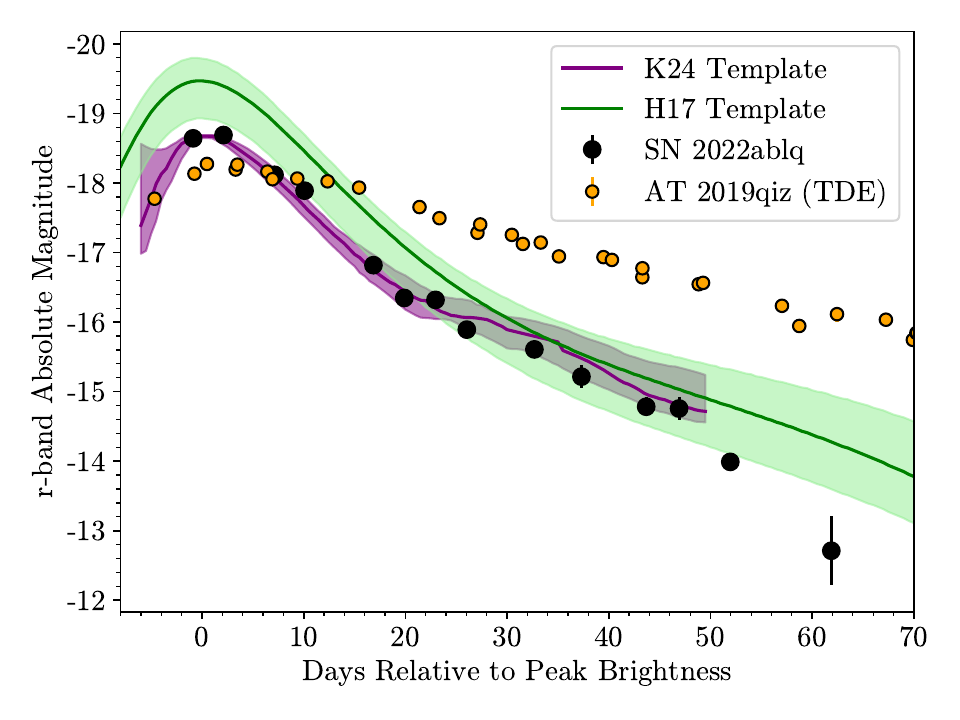}
    \includegraphics[width=0.45\textwidth]{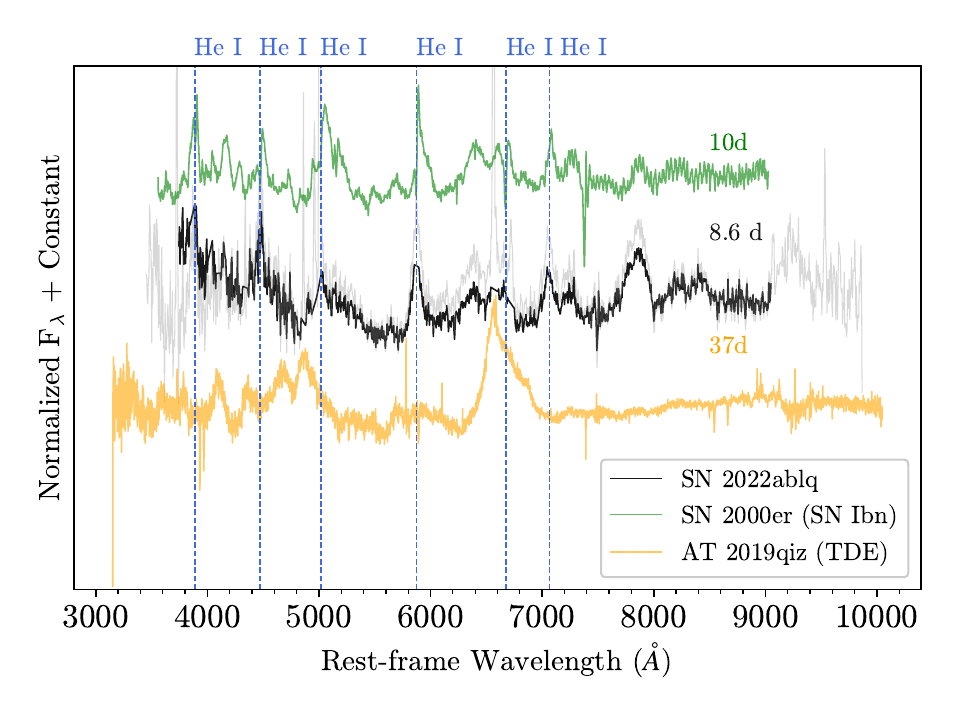}
    \caption{The photometric and spectroscopic properties of SN\,2022ablq support a Type Ibn classification. Left: The \textit{r}-band absolute magnitude light curves of SN\,2022ablq, the template SN Ibn light curves from \citet[][green]{Hosseinzadeh2017} and \citet[][purple]{Khakpash2024}, and the TDE AT\,2019qiz \citep[][orange]{Nicholl2020}. SN\,2022ablq's light curve shape is consistent with the SN Ibn templates and inconsistent with the longer-duration light curve of TDEs. Right: Optical spectra of SN\,2022ablq, Type Ibn SN\,2000er, and AT\,2019qiz. The spectrum of SN\,2022ablq before subtracting the host galaxy component is shown in light gray, for comparison with the TDE spectrum. The strong \ion{He}{1} emission features dominate the spectra of the SN Ibn but are not seen in the TDE spectrum.}
    \label{fig:comparison}
\end{figure*}

In Figure \ref{fig:comparison} we plot the \textit{r}-band absolute magnitude light curve of SN\,2022ablq. For comparison we show the averaged SN Ibn light curve behavior from \cite{Hosseinzadeh2017}, denoted as the H17 template, and \citet{Khakpash2024}, denoted as the K24 template. As the K24 template describes the light curve relative to peak brightness, we shift it to match the peak magnitude of SN\,2022ablq. Additionally we show the \textit{r}-band light curve of the TDE AT\,2019qiz \citep{Nicholl2020} as SN\,2022ablq was originally classified as a TDE \citep[][see Section \ref{subsec:classification}]{Fulton2022}. The behavior of SN\,2022ablq is consistent within the errors with both SN Ibn template light curves throughout its evolution. While SN\,2022ablq is slightly less luminous at peak than the averaged SN Ibn light curve, \citet{Hosseinzadeh2017} note that this may be an observational effect causing underluminous SNe Ibn to be missed in wide-field survey follow-up. A more detailed light curve comparison is given in Section \ref{subsec:classification}.

\subsection{Optical Spectra}\label{subsec:spec}

\begin{figure}
    \centering
    \includegraphics[width=0.475\textwidth]{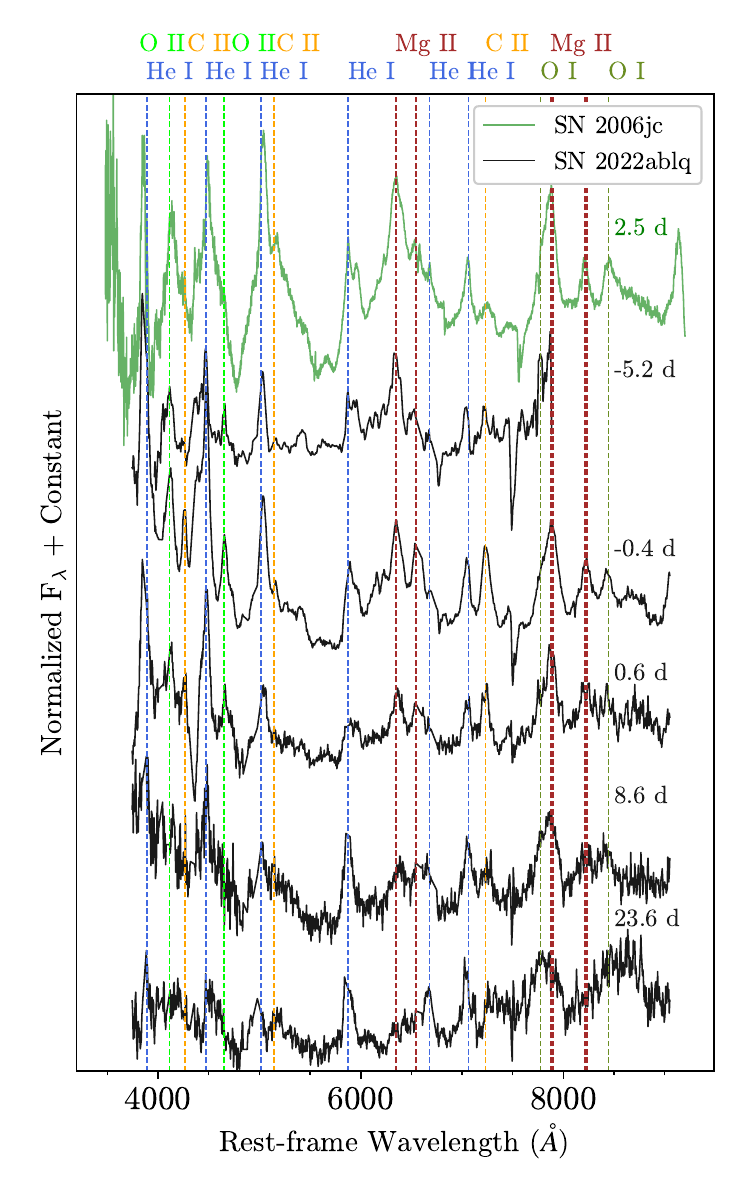}
    \caption{The normalized, host-subtracted spectral time series of SN\,2022ablq (black) compared with a representative spectrum of SN\,2006jc \citep[green;][]{Anupama2009}. Phases relative to peak optical brightness (\textit{g}-band for SN\,2022ablq and \textit{B}-band for SN\,2006jc) are given to the right of each spectrum. Notable spectral features are marked with dashed vertical lines.}
    \label{fig:specseries}
\end{figure}

Three LCO FLOYDS spectra of SN\,2022ablq were obtained between 0.6 and 23.6 days relative to peak brightness. As the SN exploded close to its host nucleus, the extracted spectra were heavily contaminated by the host continuum and narrow host emission features. To better identify the SN features, we subtract the galaxy continuum and mask these narrow lines. We use an archival SDSS spectrum \citep{SDSS2009} of the host galaxy and fit the continuum of this host galaxy spectrum with a third-order polynomial. This polynomial fit is then subtracted from the spectrum to obtain flux residuals. Using these residuals, we identify wavelength regions that are dominated by narrow features using a simple sigma cut\textemdash our algorithm best identifies emission lines to mask for wavelength regions where the residuals are more than 10$\sigma$ above the median continuum-subtracted host spectrum flux. 

Our process to remove contaminating host galaxy light from the SN spectra is as follows: we bin each scientific spectrum and the host galaxy spectrum to a common resolution, subtract the fitted host galaxy continuum, and mask the wavelength regions corresponding to strong host emission lines using our residual sigma cuts. The host-subtracted spectral time series is shown in Figure \ref{fig:specseries}. We include our NOT/ALFOSC spectrum and an additional publicly available spectrum from the Transient Name Server \citep[TNS;][]{Fulton2022,Charalampopoulos2023}. Spectral features are marked with dashed colored lines, identified by comparing noticeable features with those identified in other SN Ibn spectra \citep{Pastorello2008}. For comparison, we plot a spectrum of the prototypical Type Ibn SN\,2006jc at a similar phase. The spectral features of SN\,2022ablq closely match those of SN\,2006jc throughout the phase range probed by our observations. Strong intermediate-width \ion{He}{1} lines dominate the early-time spectra, consistent with the homogeneous behavior of SNe Ibn around peak brightness. \ion{C}{2} lines are also noticeable, although weaker than the \ion{He}{1} lines, at these phases. We also find broader emission lines of \ion{Mg}{2} as well as \ion{O}{1} and \ion{O}{2} at later phases. These lines are also consistent with those seen in SN\,2006jc \citep[e.g.,][]{Pastorello2007}.  

The constant intermediate-width \ion{He}{1} features are formed in the shell of CSM and ejecta that are swept up between the SN forward and reverse shocks, which reprocess the energetic shock photons into the UV and optical \citep{Smith2017}. Measurements of the widths of these spectral features therefore provide a lower limit on the velocity of the shocks, constraining a free parameter in the analysis of the SN X-ray emission (see Section \ref{sec:x-ray}). We measure the full-width at half-maximum, and therefore the lower bound on the shock velocity, from these features as follows. We estimate the uncertainty in our host galaxy-subtracted spectrum at +8.6 days by subtracting the observed spectrum with a smoothed spectrum using a Savitsky-Golay filter of order 3 and a 31 \AA{} window, chosen to preserve spectral features while smoothing over random noise fluctuations. We then run a Monte Carlo simulation with 3000 iterations. In each iteration, we fit a Gaussian to the \ion{He}{1} 5876\AA{} feature after varying the flux in each wavelength bin by a random value, drawn from a normal distribution centered about the calculated uncertainty at that bin. The full-width at half-maximum of the best-fit Gaussian, as well as its associated errors, give a measure of the feature width. The shock velocity is then calculated from the weighted average of each Monte Carlo width measurement.

This procedure yields an estimate of the shock velocity $v_{\text{sh}} \geq 4900 \pm 150$ km s$^{-1}$, consistent with the shock velocities estimated or assumed for other interacting SNe \citep{Immler2008,Dwarkadas2011,Chandra2012,Ofek2014}. For example, \citet{Immler2008} estimate the shock velocity of SN\,2006jc to be 9000 km s$^{-1}$, although this value may be more representative of the freely-expanding SN ejecta, rather than the swept-up shell of shocked material \citep[e.g.,][]{Pastorello2007}. The value inferred for SN\,2022ablq also places its shock in the collisionless regime for all moderate Thomson optical depths \citep[$\tau_T$ $\lesssim$ 10$^2$;][]{Margalit2022}. This information on the SN shock is critical to constraining the X-ray emission, as we discuss in Section \ref{sec:x-ray}. 

\subsection{Classification} \label{subsec:classification}

SN\,2022ablq was originally classified as a TDE on the TNS owing to its coincidence with the center of its host galaxy and contaminating host galaxy lines in its classification spectrum\textemdash an environment unlike that of SN\,2006jc, which exploded relatively offset from its host galaxy \citep{Foley2007}. \citet{Fulton2022} report spectral matches of this classification spectrum with both SNe Ibn (at +10 days relative to peak brightness) and TDEs (at +30 days relative to peak brightness). Approximately three months later, \citet{Charalampopoulos2023} reclassified the SN as a Type Ibn SN. In this section, we verify the classification of SN\,2022ablq as a Type Ibn SN on the basis of its light curve evolution, spectral features, and host galaxy properties.

The light curves of SNe Ibn and TDEs are quite different\textemdash the former are characterized by homogeneous light curve morphologies, including a short rise time (typically fewer than 10 days) and rapid decline rates \citep[typically 0.1 mag day$^{-1}$ in the optical;][]{Hosseinzadeh2017}, whereas TDEs generally rise to peak brightness over a longer time (roughly 30 days) and decay at a slower rate than SNe Ibn. In Figure \ref{fig:comparison} we show the \textit{r}-band light curve of SN\,2022ablq compared with light curves of typical SNe Ibn \citep{Hosseinzadeh2017,Khakpash2024} and the H$+$He TDE AT\,2019qiz \citep{Nicholl2020}. We find better agreement between the light curve shape, rise time, and peak brightness of the SNe Ibn than with the TDE. 

We also carefully classify an early-time spectrum of SN\,2022ablq, obtained 0.6 days after maximum light with FLOYDS. We use the SuperNova IDentification code \citep[SNID;][]{Blondin2007} with updated stripped-envelope template libraries \citep{Liu2014,Liu2016,Modjaz2016,Williamson2019,Williamson2023}. We also utilize SNID's galaxy line masking to avoid biasing our matches from the contaminating host galaxy flux. SNID returns best matches between SN\,2022ablq and the Type Ibn SN\,2000er 10 days after maximum brightness at the redshift of the host galaxy. We find no reliable matches with TDEs or other non-SN sources. Repeating this procedure with the spectrum obtained 8.6 days after maximum light again yields SN Ibn matches at similar phases. We compare this spectrum of SN\,2022ablq with the spectrum of SN\,2000er \citep{Pastorello2008} and the spectrum of the TDE AT\,2019qiz \citep{Nicholl2020} in Figure \ref{fig:comparison}. The AT\,2019qiz spectrum is at a phase $+$37 days to reproduce the match reported by \citet{Fulton2022}. The strong He I emission features observed in SN\,2022ablq are most similar to those seen in SN\,2000er and are not found in the TDE spectrum, which has broader H emission features than those found in the host-subtracted SN Ibn spectra. 

\begin{figure*}
    \centering
    \includegraphics[width=0.45\textwidth]{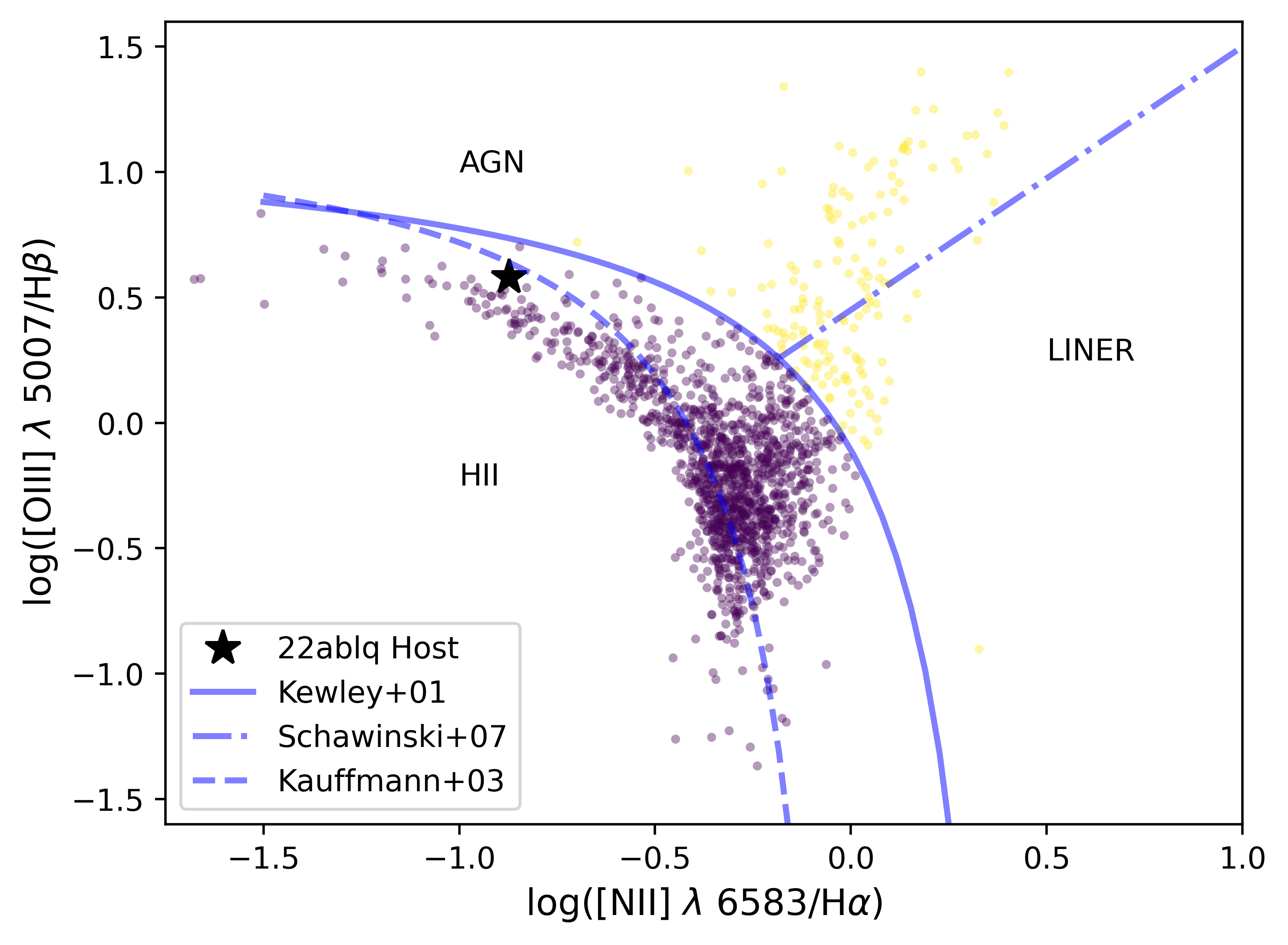}
    \includegraphics[width=0.45\textwidth]{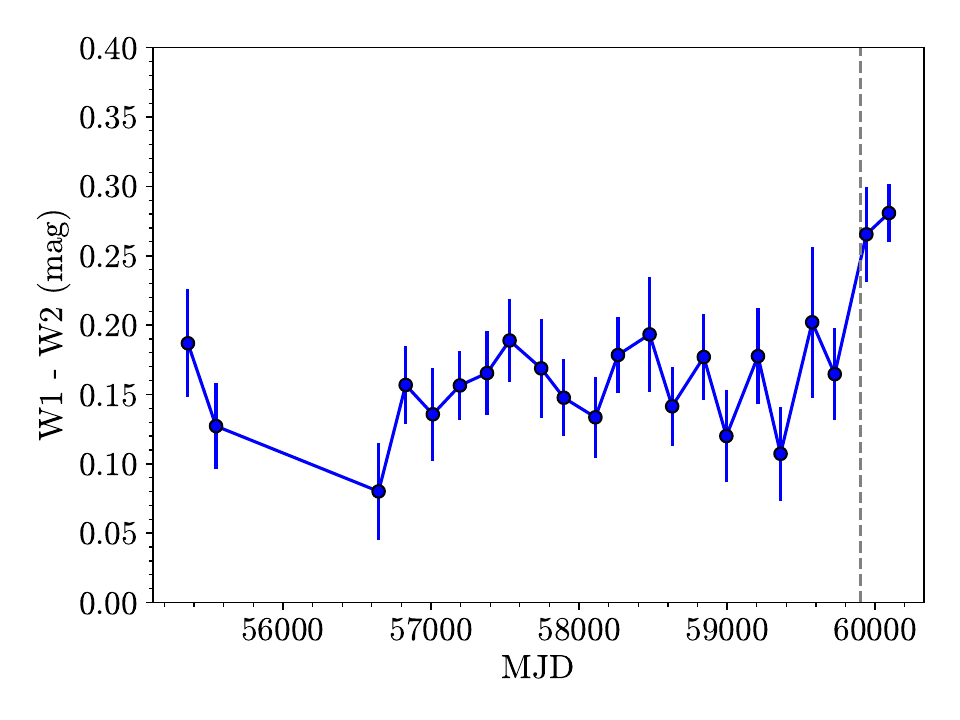}
    \caption{The host galaxy properties of SN\,2022ablq disfavor an AGN origin to the multiband emission. Left: The locus of star-forming galaxies in the parameter space of \cite{Baldwin1981} and \cite{Osterbrok1985} The properties of SN\,2022ablq are measured from an archival SDSS spectrum, compared with literature values \citep{SDSS2009} and cutoffs \citep{Kewley2001,Kauffmann2003,Schawinski2007} The host of SN\,2022ablq falls outside the cluster of AGN and LINER galaxies (yellow points) in this space, and instead is consistent with an \ion{H}{2} galaxy (purple points). Right: ALLWISE and NEOWISE \textit{W1 - W2} colors of the host of SN\,2022ablq. The explosion epoch of SN\,2022ablq is marked with a dashed gray line. The colors fall well below the AGN threshold \textit{W1 - W2} $>$ 0.8 mag during the $\sim$ 10 years before explosion.}
    \label{fig:host}
\end{figure*}

Finally, we carefully check whether any of the multi-wavelength emission we observe from SN\,2022ablq could be attributed to underlying Active Galactic Nucleus (AGN) activity. Star-forming galaxies can be classified as AGN, \ion{H}{2}, or LINER galaxies based on the relative strengths of their [\ion{O}{3}], [\ion{N}{2}], and H$\alpha$ and H$\beta$ emission lines \citep{Baldwin1981}. Following this classification schema, we measure these line strengths using an archival SDSS spectrum of the host of SN\,2022ablq. The results are plotted in Figure \ref{fig:host}, compared with a sample of SDSS galaxies \citep{SDSS2009}. We find that the host of SN\,2022ablq falls within the \ion{H}{2} locus of galaxies, separate from AGN and Seyfert galaxies.

AGN also have known color behavior in the infrared; in the Wide-field Infrared Survey Explorer \citep[WISE;][]{Wright2010} filter system, AGN have an expected infrared color \textit{W1} - \textit{W2} $>$ 0.8 mag. We utilize pre-explosion archival WISE data of the host of SN\,2022ablq to measure its temporal color behavior. Data were gathered from the ALLWISE and NEOWISE surveys from 2010 to 2023 \citep{Mainzer2011,Mainzer2014}. We filter data based on a series of quality cuts\textemdash those with frame quality $\texttt{qi\_fact} > 1$, no contamination and confusion ($\texttt{cc\_flag} = 0$) and small photometric errors ($<0.15$ mag)\textemdash and combine detections in averaged 30 day bins. These cuts leave us with 22 epochs of data, at roughly six month intervals, between June 2010 and May 2023. After binning the W1 and W2 photometry to each epoch, we calculate the infrared \textit{W1} - \textit{W2} color and display it in Figure \ref{fig:host}. The WISE photometry reveals that the host galaxy never approached the AGN threshold of \textit{W1} - \textit{W2} $>$ 0.8 during the time period we investigate, again disfavoring an AGN origin to the multi-wavelength emission of SN\,2022ablq \citep{Stern2012}.

In summary, we find that our classification of SN\,2022ablq is robust and consistent. Photometric and spectroscopic comparisons reveal similarities between SN\,2022ablq and other SNe Ibn, while analysis of pre-explosion host galaxy data disfavors an AGN origin to the multi-wavelength emission. We therefore continue with the remainder of our analysis, assuming that we are investigating the true behavior of the SN itself.

\section{X-ray Analysis}\label{sec:x-ray}

\begin{deluxetable*}{ccccc}[t!]
\tablecaption{\swift{} XRT Detections \label{tab:xrays}}
\tablehead{
\colhead{Phase Range\tablenotemark{a}} & \colhead{Exposure Time} & \colhead{Count Rate} & \colhead{Unabsorbed Flux\tablenotemark{b}} & \colhead{$<E_{avg}>$\tablenotemark{c}} \\ 
\colhead{(Days)} & \colhead{(ks)} & \colhead{(10$^{-3}$ s$^{-1}$)} & \colhead{(10$^{-14}$ erg cm$^{-2}$ s$^{-1}$)} & \colhead{(keV)}}
\startdata
15--24 & 9.6 & 0.56 $\pm$ 0.25 & 2.54 $\pm$ 1.13 & 2.3\\
38--47 & 7.9 & 1.45 $\pm$ 0.43 & 6.64 $\pm$ 1.97 & 2.3\\
51--55 & 5.2 & 1.68 $\pm$ 0.58 & 8.36 $\pm$ 2.89 & 3.5\\
66--96 & 13.9 & 0.42 $\pm$ 0.18 & 2.0 $\pm$ 0.86 & 2.9\\
97--118 & 10.3 & 0.55 $\pm$ 0.24 & 2.53 $\pm$ 1.1 & 2.5\\
349\tablenotemark{d} & 1.3 & 0.72 $\pm$ 0.77 & 3.72 $\pm$ 3.98 & 4.0\\
\enddata
\tablenotetext{a}{Phase range of stacked observations, relative to estimated explosion date}
\tablenotetext{b}{Integrated from 0.2 keV -- 10 keV assuming N$_{H}$ = 1.58$\times$10$^{21}$ }
\tablenotetext{c}{Average extracted photon energy}
\tablenotetext{d}{No significant X-ray flux was detected above the background level}
\end{deluxetable*}

SN\,2022ablq was detected in the X-rays by \swift{} XRT between 15 and 118 after explosion. To increase the signal-to-noise from these detections, we stack the XRT images using \texttt{xrtpipeline} such that each stacked image contains roughly equal counts in the aperture centered on the SN position. This process yields five combined XRT images from 20 epochs. An additional XRT observation was obtained 350 days after explosion. We find no evidence of statistically significant X-ray flux above the background measurement at this epoch.

Given the relatively low shock velocity, estimated from the intermediate-width optical spectra features and supported by the lack of broad-lined features, we assume the observed X-rays are created entirely by free-free emission from the shock-heated CSM, with negligible contributions from inverse Compton scattering due to shock-accelerated electrons \citep{Margalit2022}. To estimate the unabsorbed X-ray flux, we attempt to fit the X-ray spectrum obtained from the 0.2 -- 10 keV counts extracted from each stacked image. We use \texttt{XSELECT} to extract the spectrum from each stacked image, within the aperture described in Section \ref{subsec:swiftxrt}, and \texttt{XSPEC} to fit the observed spectrum with both thermal bremsstrahlung and power law models. However, due to the low number of counts in each image, the fitted parameters remain quite unconstrained, with low reduced $\chi^2$ values ($\ll 1$) which signify over-fitting of each model. Therefore, following the procedure of \citet{Immler2008}, we estimate the flux using the \texttt{XSPEC} bremsstrahlung model with a combined Milky Way and host H column density $N_H$ = 1.58$\times$10$^{21}$, derived from our extinction estimate in Section \ref{sec:methods} using the empirical relationship from \citet{Guver2009}, and a fixed temperature given by the average photon energy extracted from each stacked image. We use the web-based PIMMS tool\footnote{https://heasarc.gsfc.nasa.gov/cgi-bin/Tools/w3pimms/w3pimms.pl} to calculate the unabsorbed flux in this way for each stacked image, and convert this flux to an unabsorbed luminosity assuming the estimated total absorption column density as well as the distance to the host of SN\,2022ablq. These values are given in Table \ref{tab:xrays}.

\begin{figure*}
    \centering
    \includegraphics[width=0.8\textwidth]{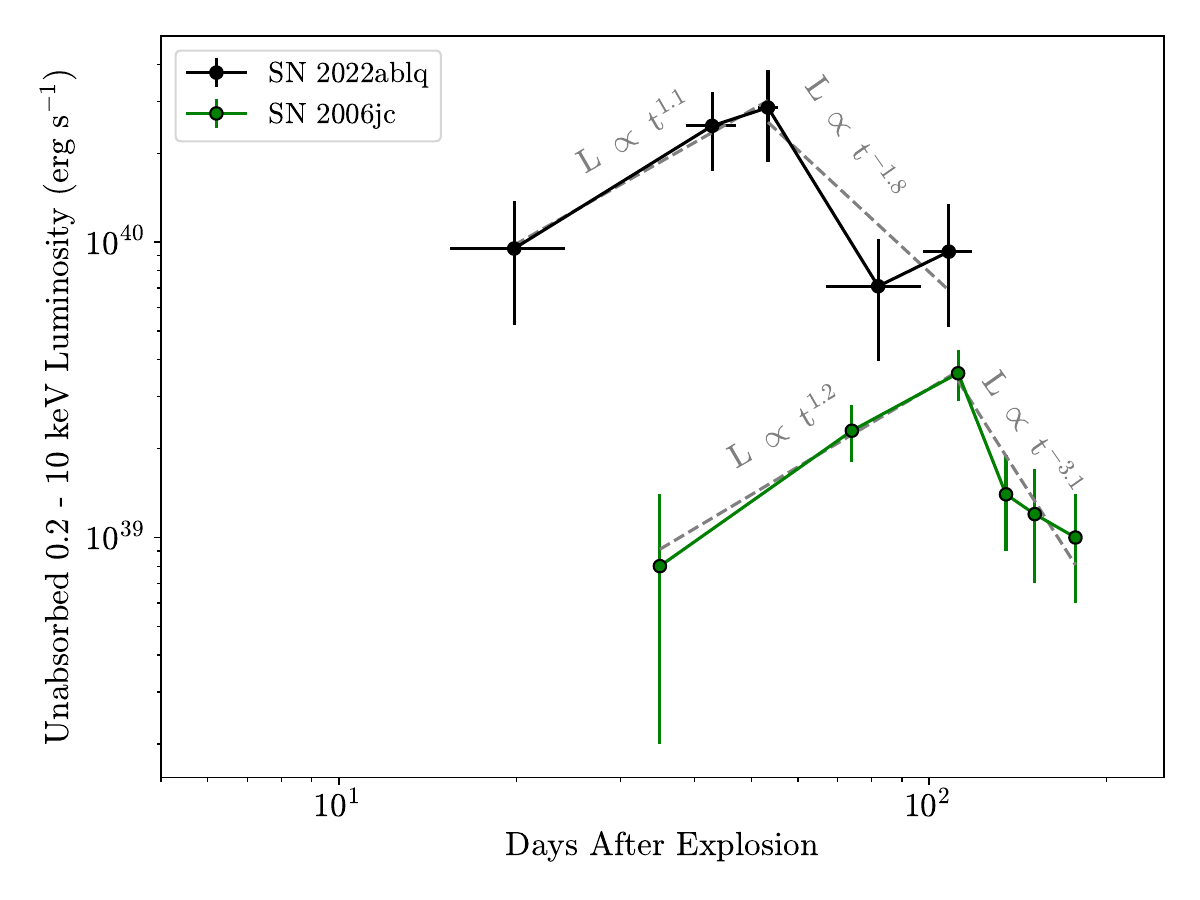}
    \caption{The 0.2 - 10 keV light curve of SN\,2022ablq (black), measured from the stacked \swift{} XRT images, compared with the X-ray light curve of the Type Ibn SN\,2006jc (green) from \citet{Immler2008}. Both X-ray light curves have qualitatively similar shapes, but the emission from SN\,2022ablq is more luminous and peaks at an earlier phase. Broken power law fits to both light curves are shown with dashed gray lines. Rising and declining portions of the light curve are fit separately. The power law indices reveal an inhomogeneous circumstellar environment for both SNe, with remarkably similar evolution during the rising portion of the light curves.}
    \label{fig:xraylc}
\end{figure*}

We present the 0.2 -- 10 keV X-ray light curve of SN\,2022ablq in Figure \ref{fig:xraylc}. For comparison, we also plot the X-ray light curve of SN\,2006jc, the only other SN Ibn with a published X-ray light curve \citep{Immler2008}. SN\,2022ablq is roughly an order of magnitude more luminous in the X-rays than SN\,2006jc and reached a peak X-ray luminosity almost 60 days before the latter, relative to their respective explosion epochs. However, both objects show similar X-ray light curve morphologies, including a sharp rise to peak followed by a rapid decay below the detection threshold of \swift{} XRT.

The light curve shapes of these SNe Ibn are qualitatively different than those of other SNe X-ray light curves. For example, SNe IIn often display long-lasting, slowly-evolving X-ray emission \citep{Chandra2018}. \citet{Immler2008} used the X-ray light curve shape of SN\,2006jc to infer the presence of a detached shell of CSM, likely produced by a precursor outburst 2 years prior to the SN explosion that was serendipitously observed in the optical \citep{Pastorello2007}. 

The time-evolution of the X-ray luminosity is a direct measure of the CSM density profile\textemdash as the shock traverses the CSM, it traces its radial distribution \citep[e.g.,][]{Dwarkadas2012}. Assuming an adiabatic shock (verified later in this Section), a steep ejecta density gradient, and a CSM density distribution described as $\rho \propto r^{-s}$, where $s$ is the CSM power law index, the X-ray luminosity is predicted to evolve as $L_X \propto t^{3 - 2s}$ \citep{Immler2001,Dwarkadas2016}. In Figure \ref{fig:xraylc} we show the piece-wise power law fits to the observed X-ray light curves of SN\,2022ablq and SN\,2006jc. Both SNe exhibit remarkably similar rising power laws ($\propto t^{1}$) until reaching peak luminosity. After peak the SNe follow slightly different power laws, with SN\,2022ablq declining as $t^{-2}$ and SN\,2006jc declining more rapidly as $t^{-3}$. This broken power law evolution indicates a complex CSM distribution, as seen in other SNe with X-ray detections \citep{Dwarkadas2012}. The observed rise may be caused entirely by decreasing absorption column density as the SN shock traverses more of the CSM. In this case, the nearby CSM provides additional absorbing material which our estimate of the column density does not account for. However, our low-resolution X-ray imaging does not allow for a study of the time-evolving column density of the nearby CSM. Another explanation is that the CSM has a shallower density profile ($\rho \propto r^{-1}$) than predicted for steady-state wind-driven mass loss ($s=2$). In either case, the rise in the X-ray emission points to dense, confined CSM that the shock is traversing at those phases. After the peak of the X-ray emission, the inferred density distribution for SN\,2022ablq scales as $r^{-2.5}$, suggesting the shock has reached the outermost part of the densest CSM and is now traversing the lower density outer region. This again is inconsistent with wind-driven mass loss, suggesting that alternative mechanisms such as binary stripping or eruptive outbursts dominate the mass-loss history during the timescale probed by these X-ray observations. Interestingly, the late-time power law slopes of both SNe are similar to that inferred by \citet{Maeda2022}, who modeled SN Ibn light curves and inferred a steep outer CSM density profile $\rho \propto r^{-3}$, which they interpreted as accelerated mass-loss leading up to the SN explosion. 

This analysis assumes the shocks are collisionless and adiabatic. Here we show that those assumptions are valid. For the observed shock emission to peak in the X-rays, the shock must be collisionless, rather than radiation-mediated \citep{Margalit2022}\textemdash in the radiation-mediated case, the X-rays are reprocessed by the dense material in the vicinity of the shock, shifting the emission into the UV and optical. Collisionless shocks may be either radiative or adiabatic, depending on the cooling timescale of the shock relative to the dynamical time. \citet{Margalit2022} provide a simple diagnostic to determine the regime of the shock from the observed peak X-ray luminosity and X-ray light curve duration, given by the ratio of the shock velocity $v_{\text{sh}}$ to a characteristic velocity

\begin{equation}
    v_{\text{rad}} = 5,200 \text{ km s}^{-1} \Big(\frac{L_X / \nu_{keV}}{10^{41} \text{erg s}^{-1}} \Big)^{1/4} \Big(\frac{t_X}{100 \text{d}}\Big)^{-1/4}
\end{equation}

where $v_{\text{sh}} > v_{\text{rad}}$ implies the shock is adiabatic. The model assumes a CSM where most of the mass $M_{\text{CSM}}$ is contained within a radius $R_{\text{CSM}}$, regardless of the exact density structure. We find that SN\,2022ablq fits these model constraints on the basis of its X-ray light curve shape, which hints at an extended shell of material. Additionally, the model assumes solar composition material. However, assuming a CSM composed entirely of He changes the derived values we present here by no more than a factor of several. From our observations SN\,2022ablq had a peak X-ray luminosity $L_X \approx 3\times 10^{40}$ erg s$^{-1}$ at an average energy $\nu \approx 2.5$ keV with duration $t_{X} \approx 100$ days, thus $v_{\text{rad}} \approx$ 2800 km s$^{-1}$. The intermediate-width optical spectral features yield an estimated shock velocity $v_{\text{sh}} \geq 4900$ km s$^{-1}$ (Section \ref{subsec:spec}). For these values $v_{\text{sh}} > v_{\text{rad}}$, therefore we are confident our assumption of adiabaticity is valid.

\begin{figure*}
    \centering
    \includegraphics[width=0.45\textwidth]{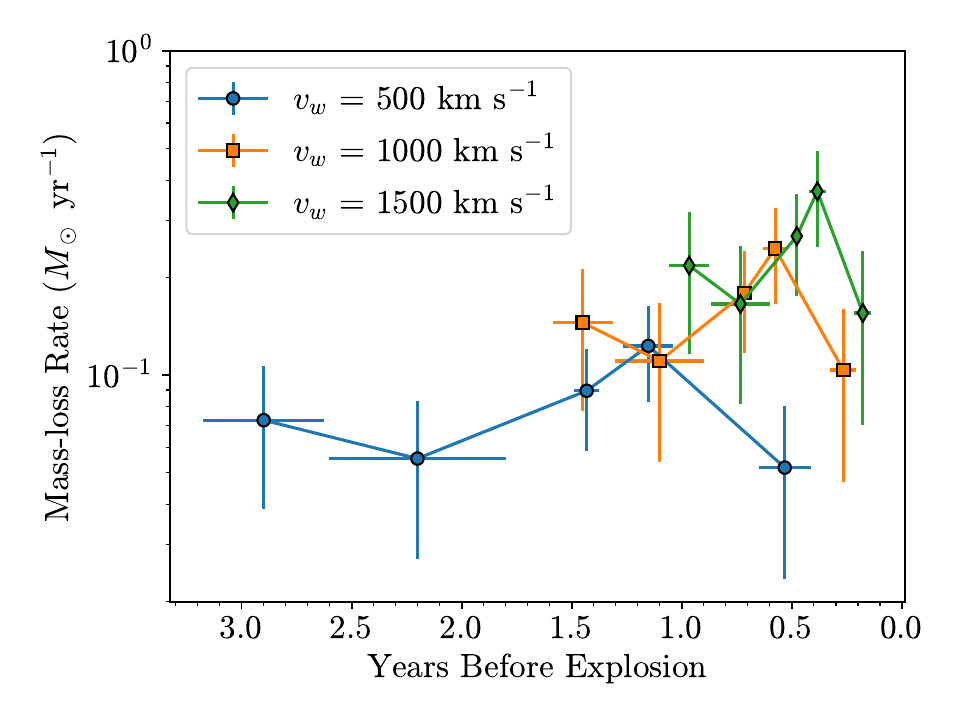}
    \includegraphics[width=0.5\textwidth]{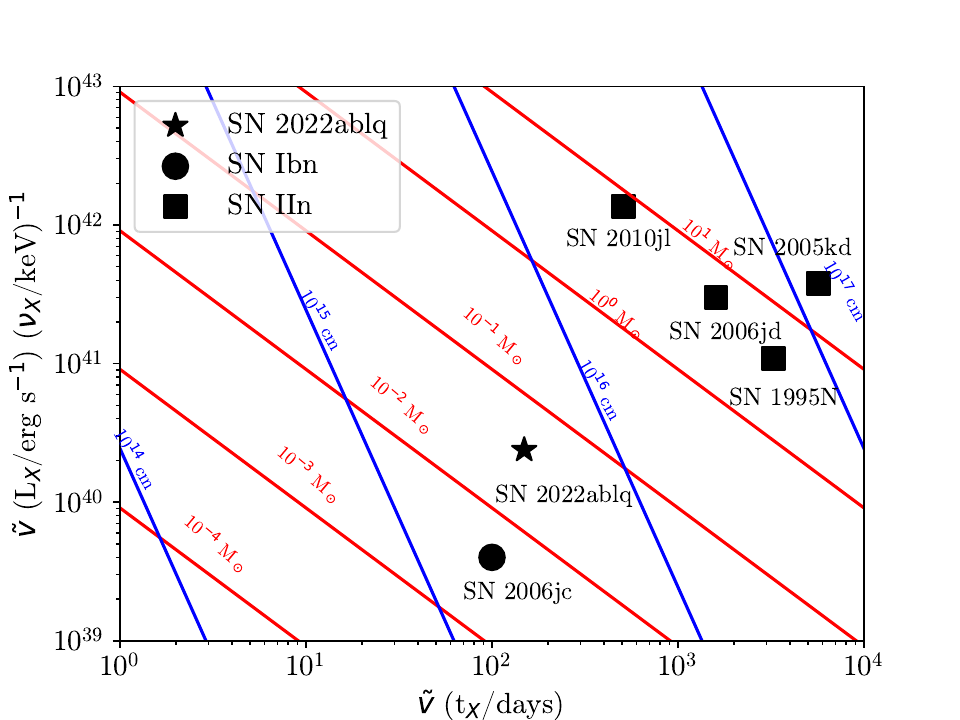}
    \caption{Left: The mass-loss rate of the progenitor of SN\,2022ablq, measured from our stacked \swift{} XRT observations, for several CSM velocities $v_w$. The data reveal enhanced mass loss, with rates between 0.05 $M_\odot$ yr$^{-1}$ and 0.5 $M_\odot$ yr$^{-1}$ during the 0.5--2 years preceding the explosion, consistent with the X-ray light curve morphology. Right: The parameter space of \citet{Margalit2022}, which relates the CSM radii (blue lines) and masses (red lines) to the X-ray light curve duration ($t_X$) and peak luminosity ($L_X / \nu_X$), scaled by $\tilde{v}$ $= v_{\text{sh}} / v_{\text{rad}}$. The CSM configuration inferred for SN\,2022ablq is marked with a black star and compared with estimate for SN\,2006jc (black circle) and several SNe IIn with well-studied X-ray light curves \citep[][black squares]{Chandra2012,Ofek2014,Chandra2015,Chandra2018}. The SNe Ibn show more confined, lower-mass CSM than the SNe IIn, potentially hinting at different mass-loss rates, timescales, and mechanisms between the progenitors of these classes, although X-ray observations of SNe Ibn at later phases are needed to probe their full CSM distributions.}
    \label{fig:csm}
\end{figure*}

In the adiabatic regime the X-ray luminosity scales with the progenitor mass-loss rate as $L_{X} \propto (\frac{\dot{M}}{v_w})^2$ \citep{Fransson1996}, where $v_w$ is the outflow velocity of the lost material. \citet{Immler2001} provide a commonly-used equation to calculate the mass-loss rate given X-ray luminosity measurements. However, as noted by \citet{Dwarkadas2012}, this dependence is only valid when the CSM profile is wind-like (and thus the luminosity scales with time as $L_{X} \propto$ $t^{-1}$), which is not true for SN\,2022ablq. Instead, we use the first principles laid out by \citet{Fransson1996} to derive the generalized mass-loss relationship between the X-ray luminosity and progenitor mass-loss rate for an arbitrary CSM profile. The unabsorbed free-free X-ray luminosity generated by the forward shock can be approximated by 
\begin{equation}
    L_{X} \approx j_{ff}(T_e)M_{\text{CSM}} \rho_{\text{CSM}} / m^2 \label{eq:masslossrate}
\end{equation}
where $j_{ff}(T_e)$ is the emissivity given by Eq. 3.8 of \citet{Fransson1996} and broadly depends on the free-free Gaunt factor, the electron temperature, and the CSM composition,
\begin{equation}
    M_{\text{CSM}} \sim \frac{\dot{M}}{v_w} \frac{R}{3-s}\big(\frac{R}{R_0}\big)^{2-s}
\end{equation}
is the mass of the CSM within a radius $R$, 
\begin{equation}
    \rho_{\text{CSM}} \sim \frac{1}{4\pi R_0^2} \frac{\dot{M}}{v_w} \big(\frac{R_0}{R}\big)^{s}
\end{equation}
is the density of the CSM at radius $R$, and $m$ is the average particle mass of the CSM. Here $R_0$ is a reference radius which we set to the radius of the shock at each X-ray epoch, estimated as the lower limit of the shock velocity multiplied by the time since explosion, allowing us to trace the mass-loss rate of the progenitor star at progressively larger radii (corresponding to earlier times before core-collapse).

From Equation \ref{eq:masslossrate} we calculate the progenitor mass-loss rate using our observed X-ray luminosities, shock velocity estimate, and CSM power law index $s$ at each epoch of X-ray measurements. We caution that the derived values are only estimates, likely within an order of magnitude, of the true progenitor mass-loss rate due to several simplifying assumptions we make in the calculations. For example, we assume a spherically symmetric CSM composed entirely of He and a constant shocked CSM temperature $T_e = 10^7$ K, although this temperature is quite unconstrained due to the low \swift{} XRT count rates (varying this value between 10$^7$K and 10$^9$K changes the inferred mass-loss rate by a factor of several). We also assume that the XRT observations from 0.2 -- 10 keV cover the entirety of the X-ray emission, which may not be the case if the shock temperature is higher than we assume. The shock velocity, inferred from the width of the optical spectral features, is also likely a lower bound and may change as the shock traverses more of the CSM. Therefore, the results of these calculations should be used as order-of-magnitude estimates of the mass-loss rate and to qualitatively infer variability in the mass-loss history.

We show our estimates of the mass-loss rate preceding SN\,2022ablq in Figure \ref{fig:csm} for different outflow velocities $v_w$. $v_w$ is normally measured from narrow emission lines in early-time spectra; however, our spectral time series begins after any such narrow lines have already vanished. Therefore we vary $v_w$ to represent the range of values measured from narrow spectral features of other SNe Ibn \citep{Pastorello2016}. Each value of $v_w$ gives a different set of times at which we measure the mass-loss rate, calculated as the ratio of the traversed radius to the outflow velocity. The mass-loss rate estimates point to a period of enhanced mass-loss of 0.05 -- 0.5 $M_\odot$ yr$^{-1}$ between $\approx$ 2.0 -- 0.5 yr before the SN explosion, depending on the exact $v_w$ value, while the later X-ray observations (probing earlier times before explosion) reveal a lower mass-loss rate. This mass-loss history is again inconsistent with steady-state stellar winds and instead implies an outburst or variable mass-loss history. Qualitatively, this is similar to the mass-loss history of SN\,2006jc, which had an observed outburst two years before the SN explosion which was also inferred from the X-ray light curve. 

As another independent probe of the CSM properties,  we use the formalism provided by \citet{Margalit2022} to estimate the total CSM mass and radius. These estimates rely on the peak X-ray luminosity $L_X$ and X-ray light curve duration $t_X$, independent of the exact CSM density profile \cite[Eqs. 45 and 46 of][]{Margalit2022}. Using our observations of SN\,2022ablq we find a CSM mass $M_{\text{CSM}} \approx 0.04$ $M_\odot$ and radius $R_{\text{CSM}} \approx 4.3\times 10^{15}$ cm. These estimates are similar to those inferred for other SNe Ibn, either through models of their bolometric light curves \citep{Pellegrino2022,Maeda2022,BenAmi2023} or similar X-ray analysis \citep{Immler2008}. We plot the distribution of CSM masses and radii inferred from X-ray observations of SNe in the \citet{Margalit2022} parameter space (Figure \ref{fig:csm}, right). While the parameter space is still sparsely populated, owing to the small number of SNe with X-ray light curves, we notice a growing trend in which SNe IIn systematically have longer-lasting and brighter X-ray emission than SNe Ibn, indicating more massive and extended CSM. Such a distribution is not unexpected\textemdash SNe IIn often have optical light curves lasting years, indicating persistent strong interaction with H-rich material possibly lost as a result of binary evolution \citep{Ercolino2024}. On the other hand, this may be a result of observational bias, as SNe IIn tend to have X-ray observations with more sensitive instruments such as \textit{Chandra} at later epochs than SNe Ibn, allowing for a more complete probe of CSM masses distributions out to larger radii. Interestingly, several ``transitional" SNe Ibn with narrow H emission have been observed, with progenitors that are suggested to be transitioning from LBV to WR stages \citep[e.g.;][]{Pastorello2015}. This X-ray parameter space serves as a novel diagnostic to determine the similarities in the circumstellar environments of SNe Ibn, SNe IIn, and these transitional objects, potentially discriminating between their possible progenitor channels. 

While SN\,2022ablq is only the second SN Ibn with unambiguous X-ray detections, it is striking how similar its X-ray and inferred CSM properties are to those of the archetypal Type Ibn SN\,2006jc. This similarity may indicate a similar progenitor to these objects, and potentially to all (X-ray luminous) SNe Ibn. While X-ray emission from SNe Ibn are rare, there has also been a lack of systematic searches for X-ray emission from these objects, particularly at the phases in which the X-ray light curves of SN\,2006jc and SN\,2022ablq peak ($\approx$ 50 -- 100 days after explosion). An analysis of all X-ray observations of SNe Ibn is beyond the scope of this work; however, such a search may elucidate the typical X-ray behavior of these transients or provide constraints on their maximum X-ray luminosities.

\section{Constraints on Pre-explosion Emission}\label{sec:preexplosiondata}

\begin{figure*}
    \centering
    \includegraphics[width=0.85\textwidth]{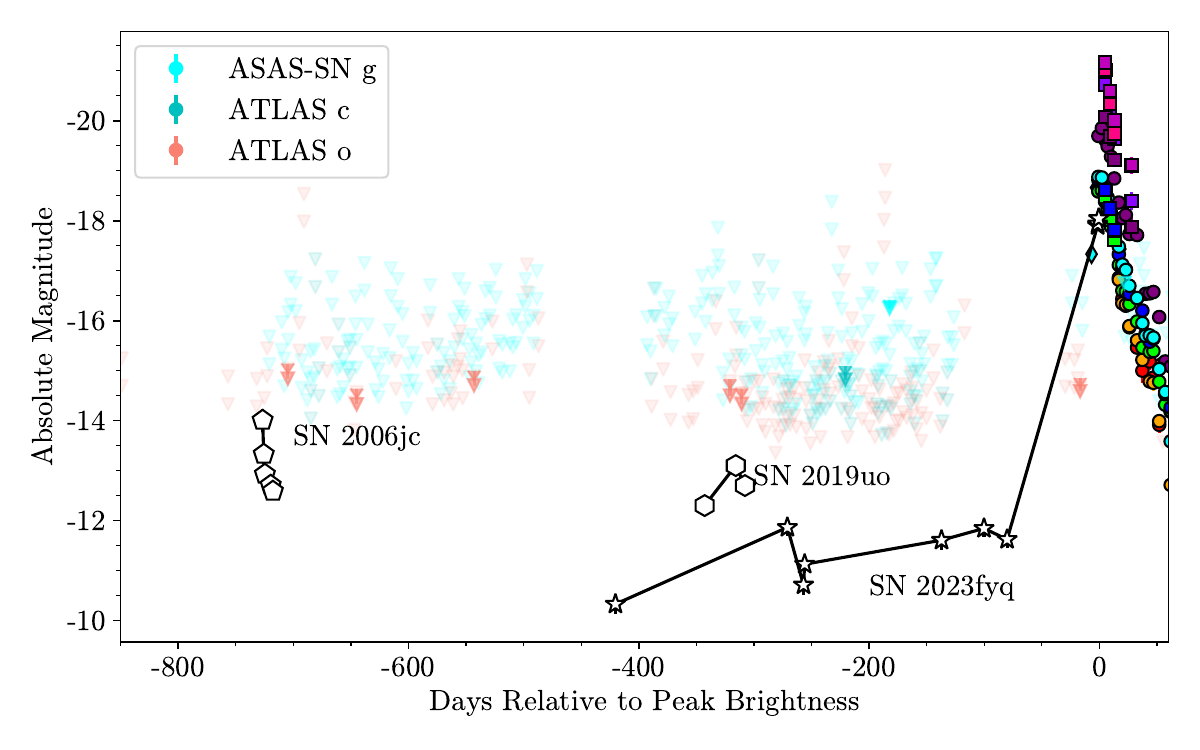}
    \caption{Pre-explosion forced photometry at the location of SN\,2022ablq from ATLAS and ASAS-SN. Individual epochs have been binned in a three day rolling window. Nondetections are shown with lighter-shaded downward-facing triangles and 3$\sigma$ detections with darker-shaded triangles. Post-explosion photometry from Figure \ref{fig:lc} is shown for reference. No significant pre-explosion emission is detected for SN\,2022ablq; the isolated 3$\sigma$ detections are more likely statistical, rather than physical, in origin. For comparison we plot \textit{r}-band precursor detections of the Type Ibn SN\,2006jc (pentagons), SN\,2019uo (hexagons), and SN\,2023fyq (stars). Our forced photometry of SN\,2022ablq rules out long-lasting precursors with absolute magnitude M $<$ -14, but are not deep enough to rule out outbursts similar to those of the other SNe Ibn.}
    \label{fig:preexp}
\end{figure*}

Observations of precursor emission to core-collapse SNe is a direct means of probing the progenitor star's enhanced mass loss during its final moments. In several occasions luminous (M $\lesssim$ -12), short-lived precursor emission has been observed in the months to years before SNe Ibn. The prototypical SN\,2006jc was preceded by a pre-explosion outburst two years before the terminal explosion \citep{Pastorello2007}. Comparing the luminosity and time scale of the X-ray emission from SN\,2006jc to this mass-loss event allowed for orthogonal constraints on the mass of its CSM and the timescale of its enhanced mass loss \citep{Immler2008}. Additionally, the Type Ibn SN\,2019uo had observed precursor emission beginning about 340 days before explosion \citep{Strotjohann2021}. More recently, extraordinarily long-lasting and variable pre-explosion emission was detected at the explosion site of the peculiar SN Ibn 2023fyq, which lasted from years to weeks before explosion \citep{Brennan2024,Dong2024}. However, it is still unclear how commonplace this pre-explosion emission is, as well as its timescales and peak luminosities; for example, pre-explosion observations of the nearby SNe 2020nxt and 2015G revealed no detectable precursor emission down to a limit M $\leq$ -14.8 and M $\leq$ -13.3, respectively \citep{Shivvers2017,Wang2024}.

We search for precursor emission to SN\,2022ablq using pre-explosion forced photometry from all-sky surveys. Given its low redshift and luminous X-ray emission\textemdash indicating strong pre-explosion mass-loss\textemdash SN\,2022ablq is a strong candidate for precursor emission detected by these surveys. Both ATLAS and ASAS-SN imaged the field containing SN\,2022ablq numerous times within the years preceding its explosion; the Zwicky Transient Facility \citep{Graham2019} also imaged the field, but did not measure any flux in forced photometry at the SN explosion site. We find that our ATLAS measurements are systematically deeper than those of ASAS-SN; therefore we focus on the former to search for faint precursor emission. We followed the analysis of \citet{Strotjohann2021}, \citet{JacobsonGalan2022}, and \citet{Rest2024} for consistency with previous studies. Our procedure is as follows. For each survey and filter combination, we first perform a 3$\sigma$ cut on intranight observations to eliminate spurious measurements. We then reject measurements with reported reduced $\chi ^2$ $\geq$ 1.4 to eliminate poor subtractions. This reduced $\chi ^2$ value was chosen to match the procedure of \citet{Strotjohann2021}. We find that this cutoff effectively reduces the number of spurious pre-explosion detections while retaining the post-explosion detections of the SN itself. 

We further remove spurious detections by extracting photometry at 8 locations in a ring-like pattern with a 16$\arcsec$ radius centered on the position of SN\,2022ablq. These ``control" light curves overlap with other regions of the host galaxy and are used to check for systematic offsets in the flux measurements; as these measurements are extracted at locations without a transient, the difference-imaged fluxes should be consistent with zero for each epoch. Any flux measurement inconsistent with zero (at the 3$\sigma$ level) indicates that the science image is of poor quality, and we discard it from our analysis. 

We repeat a similar filtering process by forcing difference-imaged photometry at the position of a nearby reference star in the Gaia catalog. Any deviation from zero flux in our forced photometry measurements likely indicates that the image subtraction was of poor quality. We discard those images with reference star flux measurements inconsistent with zero (again at the 3$\sigma$ level) from our analysis.

The above quality cuts remove approximately a third of the pre-explosion images. The remaining epochs that pass these quality cuts are then analyzed for any precursor emission. To increase the signal-to-noise of the data we bin the flux measurements using a weighted rolling average within a three day bin. This bin size is smaller than the duration of observed precursors before other SNe Ibn. We consider a 3$\sigma$ flux measurement a tentative detection and a 5$\sigma$ measurement a true detection. Additionally, we discard any 3$\sigma$ detections that are not adjacent to at least one other $\geq 3 \sigma$ detection in order to discard epochs that are likely due to statistical fluctuations or poor image quality, rather than true physical precursor emission. 

The pre-explosion photometry measurements are shown in Figure \ref{fig:preexp}. We find no conclusive evidence of precursor emission in the ATLAS or ASAS-SN pre-explosion images of SN\,2022ablq out to $\approx$800 days before explosion. Several 3$\sigma$ detections are recovered, but none are consistent across adjacent epochs, and inspecting the intranight photometry reveals these measurements are dominated by a single exposure with much higher signal-to-noise than others. We repeat this analysis with bin sizes of one and seven days but find no additional pre-explosion detections. These non-detections rule out precursor emission more luminous than $M$ $\approx$ -14 during the duty cycle of the surveys.

Interestingly, a 3$\sigma$ detection in the ATLAS \textit{o}-band is recovered approximately 9.9 days before the first ASAS-SN detection. This measurement is not adjacent to another detection at a separate epoch, nor is it high enough significance for us to deem it a true detection. However, qualitatively similar pre-explosion emission was observed for the peculiar Type Ibn SN 2023fyq, which \citet{Dong2024} concluded was evidence of enhanced mass-loss owing to either runaway binary mass transfer or final-stage core silicon burning immediately preceding the explosion. While the data of SN\,2022ablq are not significant enough to support a similar conclusion, we note that pre-explosion analyses of future nearby SNe Ibn will help constrain the fraction of precursor emission in these SNe.

\section{Comparisons with Numerical Light-curve Models} \label{sec:lcmodel}

\begin{figure*}
    \centering
    \includegraphics[width=0.65\textwidth]{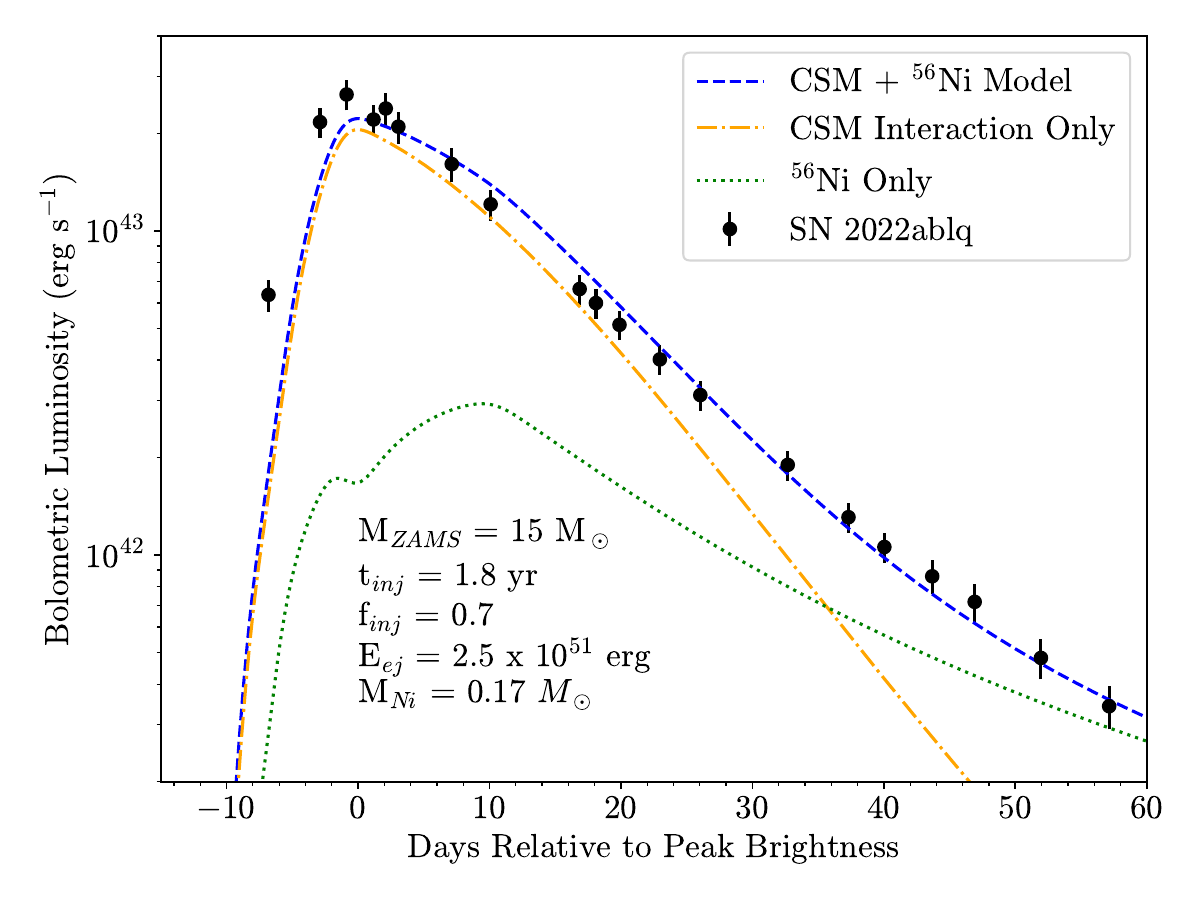}
    \caption{The pseudo-bolometric light curve of SN\,2022ablq (black points) compared to the \texttt{CHIPS} numerical light curve model. The full CSM interaction and $^{56}$Ni decay light curve is shown (blue dashed line) alongside the separate CSM (orange dashed-dotted line) and $^{56}$Ni decay (green dotted line) components. The model parameters are shown in text. There is remarkable agreement between the model light curve and our observations throughout the SN evolution, and the model parameters are in good agreement with the progenitor and mass-loss properties we infer from the X-ray detections.}\label{fig:modellcs}
\end{figure*}

\begin{deluxetable}{cccc}[t!]
\tablecaption{Pseudo-Bolometric Luminosities \label{tab:bollums}}
\tablehead{
\colhead{MJD} & \colhead{Phase} & \colhead{log$_{10}$ ($L_{\text{obs}}$)}  & \colhead{log$_{10}$ ($L_{\text{obs,err}}$)} \\
\colhead{} & \colhead{(Days)} & \colhead{(erg s$^{-1}$)} & \colhead{(erg s$^{-1}$)}}
\startdata
59907.6 & -6.8 & 42.80 & 41.85 \\
59911.5 & -2.9 & 43.34 & 42.37 \\
59913.5 & -0.9 & 43.42 & 42.45 \\
59915.6 & 1.2 & 43.34 & 42.37 \\
59916.5 & 2.1 & 43.38 & 42.44 \\
59917.5 & 3.1 & 43.32 & 42.37 \\
59921.5 & 7.1 & 43.21 & 42.28 \\
\enddata
\tablenotetext{}{This table will be made available in its entirety in machine-readable format. A portion is shown here for reference.}
\end{deluxetable}

Comparisons between the observed SN evolution and analytical or numerical models of circumstellar interaction-powered light curves give an indirect estimate of the CSM and SN properties that are independent of those inferred from X-ray observations. Recent models have been developed to physically model the post-explosion light curves of interacting SNe with pre-explosion eruptive mass loss simulations \citep{Takei2022,Takei2024}. We apply these models to our photometric observations of SN\,2022ablq to estimate the allowed range of its progenitor mass, explosion properties, and mass-loss timescale.

We begin by estimating the pseudo-bolometric luminosity of SN\,2022ablq. Our UV and optical coverage throughout the first 50 days of its evolution allows us to approximate the true bolometric luminosity, as the vast majority of SN emission peaks in the UV and optical during these phases \citep{Arcavi2022}. To do so, we utilize the code \texttt{superbol} \citep{Nicholl2018}, which calculates (pseudo-)bolometric luminosities and blackbody properties from multiband photometry by interpolating between observed photometry. We choose to use \texttt{superbol} as it inherently interpolates and extrapolates the important \swift{} photometry to match the epochs of our ground-based optical observations. To extrapolate our photometry to the epochs of \textit{g}-band observations we choose a constant-color approximation at early times, where multi-band coverage is sparse, and a fourth-order polynomial fit at late times. The observed luminosities at each epoch, integrated over the wavelength range covered by our observations, are given in Table \ref{tab:bollums}.

We then compare our pseudo-bolometric luminosity measurements to bolometric model light curves created using the open-source code \texttt{CHIPS}. \texttt{CHIPS} is a 1D radiation hydrodynamics code aimed at producing realistic CSM profiles created from eruptive outbursts in the years preceding the terminal explosions of massive stars, and has been successful in reproducing the observed light curves of SNe Ibn as well as other peculiar interacting SESNe \citep{Tsuna2023,Takei2024}. A complete description of \texttt{CHIPS} is given in \citet{Takei2022}. For the purposes of this work, we begin with the out-of-the-box SN Ibn progenitor models, which are \texttt{MESA} \citep{Paxton2011,Paxton2013,Paxton2015,Paxton2018,Paxton2019,Jermyn2023} stellar models with ZAMS masses between 15 $M_\odot$ and 29 $M_\odot$, in 1 $M_\odot$ steps, at solar metallicity. We then create different CSM profiles by varying the energy injected into the stellar envelope to produce the mass eruption, using the detached CSM models introduced in \citet{Takei2024b}. We also vary the $^{56}$Ni mass for each model to match the late-time luminosity evolution. The models were qualitatively compared by-eye to the observed pseudo-bolometric light curve to determine the most successful model parameters in reproducing the observed luminosity evolution.

The pseudo-bolometric luminosity evolution of SN\,2022ablq is compared with a well-fitting \texttt{CHIPS} model light curve in Figure \ref{fig:modellcs}. This model, which is in remarkable agreement with our observations throughout the first 70 days of the SN evolution, was produced from an $M_{\text{ZAMS}}$ = 15 $M_\odot$ star with energy equal to 0.7 of its envelope binding energy injected at $t_{\text{inj}}$ = 1.8 yr before core-collapse, producing $\sim$ 0.1 $M_\odot$ of CSM. This injection timescale is in agreement with the range of enhanced mass loss timescales that were independently estimated from the X-ray detections in Section \ref{sec:x-ray} (e.g., Figure \ref{fig:csm}), and the total CSM mass is consistent with the lower limit derived from the X-ray peak and duration. The structure of the CSM, which in this model is assumed to be shell ``detached" from the progenitor star, is also in qualitative agreement with the CSM structure evidenced by the X-ray light curve morphology, and is able to better reproduce the slow rise to peak than other models with continuous CSM profiles \citep[e.g.,][]{Takei2024}.

The $^{56}$Ni mass needed to reproduce the late-time observations in this model is higher than those inferred for many other SNe Ibn \citep[e.g.,][]{Maeda2022}. While SN\,2022ablq may have indeed produced an unusually large amount of radioactive Ni, which may be supported by the similarly-large inferred explosion energy, it is also possible that this is a result of our simplified model. For example, any asymmetries in the CSM (which cannot be captured by the 1D model here) could lead to prolonged CSM interaction, altering the late-time emission. A more prolonged mass-loss episode may also produce a more complicated CSM profile than the single detached shell we assume here. In any case, the close agreement between the observations and model predictions, particularly within the first month of explosion, is strong evidence for interaction with a detached CSM shell as the powering source for SN\,2022ablq.

\section{Discussion and Conclusions}\label{sec:discussion}

In this section we discuss constraints on the progenitor channel and progenitor mass-loss mechanism of SN\,2022ablq from our combined X-ray observations, pre-explosion forced photometry, and light curve modeling.

\subsection{Progenitor Constraints from X-ray Observations}

The mass-loss rates, CSM density distributions, and total CSM masses and radii inferred from the X-ray light curves of SN\,2022ablq and SN\,2006jc allow us to constrain their potential progenitors, and thus the potential progenitor channels of the class of SNe Ibn. WR stars are commonly thought to be the progenitors of these SNe, as their strong and fast stellar winds are a natural mechanism to strip the outer H-rich (and some of the He-rich) material \citep[e.g.,][]{Foley2007}. However, growing samples of SNe Ibn appear to be at odds with WR stars as the sole progenitor channel. Galactic WR stars have observed mass-loss rates $\dot{M} \approx 10^{-5}$ -- $10^{-4}$ $M_\odot$ yr $^{-1}$ \citep{Crowther2007}\textemdash orders of magnitude lower than the inferred mass-loss rate for the progenitors of SN\,2006jc and SN\,2022ablq during their final few years. Additionally, the variable mass-loss rates of both SN\,2006jc and SN\,2022ablq, along with their complex CSM density profiles, are inconsistent with steady-state wind-driven mass loss. 

Another possible progenitor channel is the core-collapse of a star that loses mass through episodic mass-loss episodes, either owing to late-stage nuclear burning or to interaction with a binary companion. Eruptive outbursts are a natural way to explain the CSM distribution and variable mass-loss rate. The mass-loss rates inferred for the progenitor of SN\,2022ablq are more similar to those of LBV eruptions \citep{Smith2017}. However, the composition and total mass of the CSM characteristic of these two classes are in tension, with LBVs producing H-rich CSM with masses on the order of 10$M_\odot$ \citep{Smith2006}. Additionally, while an eruption during an LBV-like phase of WR evolution has been proposed for SN\,2006jc \citep{Pastorello2007}, such a phase has not been observed directly in Galactic WR stars. 

Eruptive outbursts in the case of a H-poor progenitor may instead be evidence of instabilities owing to the onset of late-stage nuclear burning \citep{Quataert2012}. For example, \citet{Wu2021} investigated wave-driven mass loss from the onset of O/Ne and Si burning in massive stars and found that in several mass regimes (M$_{\text{ZAMS}} \lesssim 12 M_\odot$ and M$_{\text{ZAMS}} \gtrsim 30$ $M_\odot$) burning in convective regions can produce luminous ($\approx 10^{47}$ -- $10^{48}$ erg) outbursts. However, the timescales of these outbursts are more than 10 years before core collapse in the low-mass regime, and fewer than 0.1 years before core collapse in the high-mass regime. Both are broadly inconsistent with the period of accelerated mass-loss ending roughly 1 year before explosion inferred from the X-ray light curve of SN\,2022ablq, and the latter may be inconsistent with the lack of precursor emission detected during the $\sim$ month before explosion.

An alternative model that may successfully reproduce the mass-loss timescale as well as the CSM properties of SN\,2022ablq is if its progenitor were a low-mass He star with a binary companion. \citet{Wu2022} modeled the evolution of a $M_{\text{ZAMS}} \approx 15 M_\odot$ star with a neutron star companion that evolves to a He star of mass 2.5 -- 3 $M_\odot$ at the time of core O/Ne burning and found that in certain cases, the primary star overflows its Roche lobe as He-shell burning causes its envelope to rapidly expand. The primary stars in these models underwent sudden, elevated mass loss with rates $\gtrsim$ 10$^{-2}$ $M_\odot$ yr $^{-1}$, producing a range of CSM masses ($10^{-2}$ -- 1 $M_\odot$) and radii (10$^{14}$ -- 10$^{17}$ cm). These parameters are broadly consistent with the CSM properties inferred for SN\,2022ablq from its X-ray emission, and the progenitor masses in these simulations agree with the low mass (15 $M_\odot$) progenitor in our \texttt{CHIPS} model. One caveat, however, is that \texttt{CHIPS} does not model the CSM distribution from Roche-lobe overflow, which would create a more complicated density profile and CSM with significant deviations from spherical symmetry than captured by our models. This discrepancy may result in differences in our mass-loss rate and CSM mass estimates; therefore we again caution that these values should be interpreted as order-of-magnitude estimates. Further work may determine whether the Roche-lobe overflow model, which appears promising in its ability to reproduce the complicated mass-loss history of SNe Ibn, can also reproduce the observed characteristics of their explosions and the range of observed precursor luminosities. 

\subsection{Constraints on Mass-loss Mechanism from Pre-explosion Photometry and Light-curve Modeling}

Observations of pre-explosion emission are complementary to X-ray observations in determining the mass-loss histories of core-collapse SN progenitors. For SN\,2022ablq, pre-explosion forced photometry at the SN location yielded no statistically significant precursor events. Our deepest non-detections rule out precursor emission with absolute magnitudes $M \leq -14$. However, these limits are not deep enough to rule out precursors similar to those seen in three other nearby SNe Ibn. If the progenitors of SNe Ibn have similar pre-explosion mass loss events, then such an event cannot be ruled out for the progenitor of SN\,2022ablq.

Our numerical light curve modeling with \texttt{CHIPS}, which assumes an eruptive mass-loss episode, favors heightened mass-loss roughly 1.8 years before explosion. While this time frame is probed by our pre-explosion images, \texttt{CHIPS} predicts these mass-loss episodes to peak around absolute magnitude $M \approx$ -14 with a short ($\sim$ hour timescale) duration \citep{Takei2024}. Therefore, if the \texttt{CHIPS} model accurately reproduces the true precursor emission of interacting SNe, then it was likely too faint and fast-evolving to be detected by our forced photometry measurements. 

The strong agreement between our post-explosion photometry and the numerical \texttt{CHIPS} light curve gives support for an eruptive mass-loss episode, independent of our analysis of the X-ray emission. The CSM density produced in this model is inconsistent with an $r^{-2}$ profile expected from a steady-state stellar wind. This, along with the variable X-ray light curve and enhanced mass-loss rate inferred from our X-ray analysis, leads us to disfavor stellar winds from massive (e.g., WR) stars and instead to favor highly variable outbursts or eruptions as the mass-loss mechanism producing the CSM around the progenitor of SN\,2022ablq, and potentially other SNe Ibn.

\subsection{Concluding Remarks}

We have presented multi-wavelength observations of SN\,2022ablq, a nearby Type Ibn SN. SN\,2022ablq is only the second SN Ibn with X-ray detections, offering a remarkable opportunity to directly probe its pre-explosion mass loss history and progenitor channel. An analysis of these observations leads to the following conclusions: 

\begin{itemize}
    \item SN\,2022ablq is a typical Type Ibn SN photometrically and spectroscopically, with a similar optical peak magnitude, decline rate, and spectroscopic features to other members of its class;
    \item \swift{} XRT observations reveal luminous X-ray emission, peaking at 3$\times$10$^{40}$ erg s$^{-1}$, likely stemming from the ejecta-CSM shock front, with a qualitatively similar evolution to the X-ray emission of the prototypical Type Ibn SN\,2006jc;
    \item From the X-ray emission we derive mass-loss rates between 0.05 $M_\odot$ yr$^{-1}$ and 0.5 $M_\odot$ yr$^{-1}$ during the 0.5 -- 2 years before explosion, assuming spherical symmetry, with elevated mass-loss rates immediately preceding core-collapse;
    \item An analysis of the X-ray emission yields an estimate of the CSM mass $M_{\text{CSM}}$ $\geq$ 0.04 $M_\odot$ and radius $R_{\text{CSM}}$ $\geq$ 4 $\times$ 10$^{15}$ cm, similar to estimates for SN\,2006jc but smaller than those for other interaction-powered SNe such as Type IIn SNe;
    \item No pre-explosion emission is detected in forced photometry down to absolute magnitude $M \leq -14$, which is not deep enough to rule out a similar precursor event to those observed in the Type Ibn SNe\,2006jc, 2019uo, and 2023fyq;
    \item The bolometric light curve is well fit by numerical models of a 15 $M_\odot$ ZAMS progenitor star which lost a significant fraction of its He-rich envelope via an eruptive outburst 1.8 years before core-collapse, consistent with the mass-loss timescale and CSM geometry inferred from the X-ray detections.
\end{itemize}

These conclusions collectively disfavor a massive Wolf-Rayet progenitor and instead point to a lower-mass ($\lesssim$ 25 $M_\odot$) progenitor star which underwent variable, enhanced mass loss in its final $\sim$ year. The timescale and rate of this mass loss is inconsistent with stellar winds and instead suggests an eruptive outburst, Roche-lobe overflow, or other exotic mass-loss mechanism as the main mechanism producing the CSM surrounding the progenitor of SN\,2022ablq. In particular, the CSM properties we infer, including the mass and radius of the CSM, are more similar to those properties estimated for the Type Ibn SN\,2006jc and systematically lower than those estimated for Type IIn SNe. This may suggest separate mass-loss mechanisms and possibly progenitor channels for these two classes of objects.

X-ray detections of Type Ibn SNe are exceedingly rare--16 years separated the detections of SN\,2006jc and SN\,2022ablq in the X-rays. However, this study demonstrates the utility of X-ray observations in revealing the progenitor channels of stripped-envelope SNe and the mass-loss mechanisms of their progenitor stars. In the future, targeted follow-up of interacting, stripped-envelope SNe such as Type Ibn SNe with more sensitive instruments such as \textit{Chandra} \citep{Weisskopf2000} and at higher energies with instruments such as \textit{NuSTAR} \citep{NuSTAR2013} will be needed to probe the CSM properties of these objects at larger redshifts and longer phases. Additionally, the advent of the Legacy Survey of Space and Time at the Rubin Observatory \citep{Ivezic2019} will greatly increase the number of detections of precursor emission preceding core-collapse SNe, allowing us to tightly constrain the rates of such events for the first time. This in turn will allow for a better understanding of the poorly-understood end stages of massive star evolution, including the mechanisms responsible for their heightened mass loss recently evidenced by observations of core-collapse SNe \citep[e.g.,][]{Hiramatsu2023,Brennan2024,Dong2024}\textemdash an open question in stellar evolution.

\begin{acknowledgments}
C.P, M.M., and R.B.W acknowledge support in part from ADAP program grant No. 80NSSC24K0180 and from NSF grant AST-2206657. This work makes use of observations from the Las Cumbres Observatory global telescope network. The LCO group is supported by NSF grants AST-1911151 and AST-1911225. We acknowledge the use of public data from the Swift data archive. D.T. is supported by the Sherman Fairchild Postdoctoral Fellowship at the California Institute of Technology. Research by Y.D. is supported by NSF grant AST-2008108. K.A.B. is supported by an LSST-DA Catalyst Fellowship; this publication was thus made possible through the support of Grant 62192 from the John Templeton Foundation to LSST-DA.  P.C. acknowledges support via Research Council of Finland (grant 340613).

\end{acknowledgments}

\bibliography{references}

\software{This work made use of the following software packages: \texttt{astroML} \citep{astroML}, \texttt{astropy} \citep{astropy:2013, astropy:2018, astropy:2022}, \texttt{CHIPS} \citep{Takei2022,Takei2024}, \texttt{HEAsoft} \citep{HEAsoft}, \texttt{HOTPANTS} \citep{Becker2015}, \texttt{jupyter} \citep{2007CSE.....9c..21P, kluyver2016jupyter}, \texttt{matplotlib } \citep{Hunter:2007}, \texttt{numpy } \citep{numpy}, \texttt{pandas } \citep{mckinney-proc-scipy-2010, pandas_10957263}, \texttt{python } \citep{python} and \texttt{scipy } \citep{2020SciPy-NMeth, scipy_10840233}.  Software citation information aggregated using \textit{The Software Citation Station} \citep{software_citation_station}.}

\end{document}